\pdfoutput=1
\documentclass{aa}
\usepackage{epstopdf}
\usepackage[varg]{txfonts}
\usepackage{natbib}
\bibpunct{(}{)}{;}{a}{}{,} 
\usepackage{hyperref}

\begin{document}

\title{Synchrotron Cooling in Energetic Gamma-Ray Bursts Observed by the \textit{Fermi} Gamma-Ray Burst Monitor}

\author{Hoi-Fung Yu\inst{\ref{inst1},\ref{inst2}} \and Jochen Greiner\inst{\ref{inst1},\ref{inst2}} \and Hendrik van Eerten\inst{\ref{inst1}}\thanks{Fellow of the Alexander v. Humboldt Foundation} \and J. Michael Burgess\inst{\ref{inst3},\ref{inst4}} \and P. Narayana Bhat\inst{\ref{inst5}} \and Michael S. Briggs\inst{\ref{inst5}} \and Valerie Connaughton\inst{\ref{inst5}} \and Roland Diehl\inst{\ref{inst1}} \and Adam Goldstein\inst{\ref{inst6}} \and David Gruber\inst{\ref{inst7}} \and Peter A. Jenke\inst{\ref{inst5}} \and Andreas von Kienlin\inst{\ref{inst1}} \and Chryssa Kouveliotou\inst{\ref{inst6}} \and William S. Paciesas\inst{\ref{inst8}} \and V{\'e}ronique Pelassa\inst{\ref{inst5}} \and Robert D. Preece\inst{\ref{inst5},\ref{inst9}} \and Oliver J. Roberts\inst{\ref{inst10}} \and Bin-Bin Zhang\inst{\ref{inst5}}}

\institute{Max-Planck-Institut f{\"u}r extraterrestrische Physik, Giessenbachstra{\ss}e 1, 85748 Garching, Germany\\ \email{sptfung@mpe.mpg.de}\label{inst1}
\and Excellence Cluster Universe, Technische Universit{\"a}t M{\"u}nchen, Boltzmannstra{\ss}e 2, 85748 Garching, Germany\label{inst2}
\and The Oskar Klein Centre for Cosmoparticle Physics, AlbaNova, SE-106 91 Stockholm, Sweden\label{inst3}
\and Department of Physics, KTH Royal Institute of Technology, AlbaNova, SE-106 91 Stockholm, Sweden\label{inst4}
\and Center for Space Plasma and Aeronomic Research (CSPAR), University of Alabama in Huntsville, 320 Sparkman Drive, Huntsville, AL 35805, USA\label{inst5}
\and Astrophysics Office, ZP12, NASA/Marshall Space Flight Center, Huntsville, AL 35812, USA\label{inst6}
\and Planetarium S\"{u}dtirol, Gummer 5, 39053 Karneid, Italy\label{inst7}
\and Universities Space Research Association, 320 Sparkman Drive, Huntsville, AL 35805, USA\label{inst8}
\and Department of Space Science, University of Alabama in Huntsville, 320 Sparkman Drive, Huntsville, AL 35899, USA\label{inst9}
\and School of Physics, University College Dublin, Belfield, Dublin 4, Ireland\label{inst10}
}

\abstract
{In this paper we study the time-resolved spectral properties of energetic gamma-ray bursts (GRBs) with good high-energy photon statistics observed by the Gamma-Ray Burst Monitor (GBM) onboard the \textit{Fermi} Gamma-Ray Space Telescope.}
{To constrain in detail the spectral properties of GRB prompt emission on a time-resolved basis and to discuss the theoretical implications of the fitting results in the context of various prompt emission models.}
{Our sample comprises eight GRBs observed by \textit{Fermi} GBM in its first five years of mission, with 1~keV - 1~MeV fluence $f>1.0\times10^{-4}$~erg~cm$^{-2}$ and signal-to-noise level $\text{S/N}\geq10.0$ above 900~keV. We perform time-resolved spectral analysis using a variable temporal binning technique according to optimal S/N criteria, resulting in a total of 299 time-resolved spectra. We fit the Band function to all spectra and obtain the distributions for the low-energy power-law index $\alpha$, the high-energy power-law index $\beta$, the peak energy in the observed $\nu F_\nu$ spectrum $E_\text{p}$, and the difference between the low- and high-energy power-law indices $\Delta s=\alpha-\beta$. We also apply a physically motivated synchrotron model, which is a triple power-law with constrained power-law indices and a blackbody component, to test for consistency with a synchrotron origin for the prompt emission and obtain the distributions for the two break energies $E_\text{b,1}$ and $E_\text{b,2}$, the middle segment power-law index $\beta$, and the Planck function temperature $kT$.}
{The Band function parameter distributions are $\alpha=-0.73^{+0.16}_{-0.21}$, $\beta=-2.13^{+0.28}_{-0.56}$, $E_\text{p}=374.4^{+307.3}_{-187.7}$~keV ($\log_{10}E_\text{p}=2.57^{+0.26}_{-0.30}$), and $\Delta s=1.38^{+0.54}_{-0.31}$, with average errors $\sigma_\alpha\sim0.1$, $\sigma_\beta\sim0.2$, and $\sigma_{E_\text{p}}\sim0.1E_\text{p}$. Using the distributions of $\Delta s$ and $\beta$, the electron population index $p$ is found to be consistent with the "moderately fast" scenario which fast- and slow-cooling scenarios cannot be distinguished. The physically motivated synchrotron fitting function parameter distributions are $E_\text{b,1}=129.6^{+132.2}_{-32.4}$~keV, $E_\text{b,2}=631.4^{+582.6}_{-309.6}$~keV, $\beta=-1.72^{+0.48}_{-0.25}$, and $kT=10.4^{+4.9}_{-3.7}$~keV, with average errors $\sigma_\beta\sim0.2$, $\sigma_{E_\text{b,1}}\sim0.1E_\text{b,1}$, $\sigma_{E_\text{b,2}}\sim0.4E_\text{b,2}$, and $\sigma_{kT}\sim0.1kT$. This synchrotron function requires the synchrotron injection and cooling break (i.e., $E_\text{min}$ and $E_\text{cool}$) to be close to each other within a factor of ten, often in addition to a Planck function.}
{A synchrotron model is found consistent with the majority of time-resolved spectra for eight energetic \textit{Fermi} GBM bursts with good high-energy photon statistics, as long as both the cooling and injection break are included and the leftmost spectral slope is lifted either by inclusion of a thermal component or when an evolving magnetic field is accounted for.}
{}


\titlerunning{Synchrotron Cooling in Energetic GRBs Observed by the \textit{Fermi} GBM}
\maketitle

\section{Introduction}
\label{sect:intro}

\begin{table*}[!htbp]
\caption{The names, GBM trigger numbers, durations, fluence, detectors used, and optimal S/N for the eight bursts studied in this paper.}
\label{tab:tab1}
\centering
\begin{tabular}{llccccc}
\hline\hline
GRB Name & GBM Trigger \# & $T_{90}$ & $f$(1~keV - 1~MeV) & NaI & BGO & S/N \\
 & & (s) & ($10^{-4}\text{erg/cm}^2$) & & & \\
\hline
090902B & 090902.462 & 138.2$\pm$3.2 & 2.22$\pm$0.003 & n0, n1, n9 & b0 & 50 \\
100724B & 100724.029 & 114.7$\pm$3.2 & 2.17$\pm$0.006 & n0, n1, n2 & b0 & 40 \\
100826A & 100826.957 & 85.0$\pm$0.7 & 1.64$\pm$0.010 & n7, n8 & b1 & 30 \\
101123A & 101123.952 & 103.9$\pm$0.7 & 1.13$\pm$0.001 & n9, na & b1 & 30 \\
120526A & 120526.303 & 43.6$\pm$1.0 & 1.16$\pm$0.002 & n4 & b0 & 20 \\
130427A & 130427.324 & 138.2$\pm$3.2 & 24.62$\pm$0.012 & n6, na & b1 & 20 \\
130504C & 130504.978 & 73.2$\pm$2.1 & 1.29$\pm$0.002 & n2, n9 & b0 & 30 \\
130606B & 130606.497 & 52.2$\pm$0.7 & 2.01$\pm$0.002 & n7, n8, nb & b1 & 40 \\
\hline
\end{tabular}
\end{table*}

Gamma-ray bursts (GRBs) are the most luminous explosions in the Universe known to-date. The first GRB was discovered in 1967 \citep{Klebesadel73a}, and after over 45 years of research efforts it is now believed that GRBs originate from highly relativistic outflows from central compact sources at cosmological distances with bulk Lorentz factors $\Gamma>100$ \citep[e.g.][]{Lithwick01a,Hascoet12a}. This is often understood in terms of the "fireball model" \citep{Goodman86a,Paczynski86a,Rees92a,Rees94a,Piran99a}, where the GRB itself is produced by dissipation of kinetic energy from the relativistic flow. However, the shape of GRB spectra does not naturally fit the synchrotron spectra predicted by this model. Even after many GRB dedicated missions, e.g. the Burst And Transient Source Explorer \citep[BATSE,][]{Fishman89a,Meegan92a} onboard the \textit{Compton} Gamma-Ray Observatory (\textit{CGRO}), the \textit{BeppoSAX} satellite \citep{Boella97a}, the \textit{Swift} satellite \citep{Gehrels04a}, and the \textit{Fermi} Gamma-Ray Space Telescope \citep{Atwood09a}, no single consensus theory has emerged explaining all the features of the prompt emission, although various possibilities aside from the basic fireball model have been raised \citep[see, e.g.,][for a recent overview]{Zhang14a}.

To study the physical properties of GRB prompt emission, the observed $\gamma$-ray spectrum is usually fitted to a chosen model (either physical or empirical). Then the best fit parameters can be compared to the physical parameters used in theoretical models and computer simulations. Over the past 20 years the preferred fitting model has been the empirical Band function \citep{Band93a}, which consists of a smoothly joined broken power-law with low-energy power-law index $\alpha$, high-energy power-law index $\beta$, and a characteristic energy $E_\text{p}$ parameterized as the peak energy in the observed $\nu F_\nu$ spectrum.

Since the observed spectral behaviour varies from burst to burst and over time within a single burst, it is crucial to study the fitted parameters from a carefully selected sample of GRBs in a systematic way. Well-constrained spectral parameters are also important to distinguish among various theoretical models. However, due to the observed high-energy cutoff nature of the spectrum and the fact that it is harder to detect high-energy $\gamma$-ray photons, the high-energy power-law index is often poorly constrained for most bursts. Thanks to the broad spectral coverage of the Gamma-Ray Burst Monitor \citep[GBM,][]{Bissaldi09a,Meegan09a} onboard \textit{Fermi}, we are now able to obtain the spectral indices with good precision.

Motivated by the fact that most catalog studies of large GRB samples do not consider the quality of high-energy photon statistics \citep[e.g.,][]{Kaneko06a,Nava11a,Goldstein12a,Goldstein13a,Gruber14a,Yu14a}, we present time-resolved spectroscopy for eight energetic GRBs with good high-energy statistics in the GBM GRB zoo \citep{Bissaldi11a} to obtain an accurate measurement of $\beta$. We describe the selection criteria, analysis procedures and empirical fitting models in Sect.~\ref{sect:ana}. The observational results are presented in Sect.~\ref{sect:obs}. We present the fitting results from the standard Band function in Sect.~\ref{subsect:bandfit}, and a test synchrotron model in Sect.~\ref{subsect:syncfit}. In Sect.~\ref{sect:dis} we discuss the theoretical implications of the observed parameter distributions in the context of different models. The conclusion is given in Sect.~\ref{sect:conc}. Unless otherwise stated, all errors reported in this paper are given at the 1-$\sigma$ confidence level.

\section{GBM Data Analysis}
\label{sect:ana}

\subsection{Instrumentation}
\label{subsect:inst}

GBM is a sensitive scintillation array onboard the \textit{Fermi} satellite. It consists of twelve thallium activated sodium iodide (NaI(Tl)) detectors covering energy from 8~keV to 1~MeV and two bismuth germanate (BGO) detectors covering energy from 200~keV to 40~MeV. This provides spectral coverage over three orders of magnitude, which makes GBM a powerful observing instrument for GRB prompt emission.

\subsection{Burst, Detector, and Data Selection}
\label{subsect:data}

The sample presented in this paper are among the most energetic bursts observed by \textit{Fermi} GBM until 21 August 2013. They were selected according to two criteria: (1) total fluence in 1~keV - 1~MeV, $f>1.0\times10^{-4}$~erg~cm$^{-2}$; and (2) signal-to-noise level, $\text{S/N}\geq10.0$ above $900$~keV (i.e. the NaI limit) in the BGO. The advantage of analysing bursts having significant photon statistics above 900~keV is that the high-energy power-law index can be better constrained. Moreover, high fluence provides more statistics for time-resolved spectral analysis. Table~\ref{tab:tab1} lists the eight long GRBs (time in which 90\% of burst fluence observed, $T_{90}>2$~s) satisfying the above selection criteria. There are no short bursts in the sample because they do not satisfy our fluence criterion. GRB 130427A is the brightest burst observed by GBM. This brightness caused a pulse pile-up effect in the detectors in its complex-shaped main pulse after $t=T_0+2.4$~s. However, it also has a bright first pulse that is well suited for testing the synchrotron model \citep{Preece14a} and that satisfies our selection criteria by itself. Therefore, this first pulse ($t<T_0+2.4$~s) is included in our analysis.

For each burst, up to three NaI detectors with viewing angle less than 60 degrees and the BGO without blockage by either the Large Area Telescope \citep{Atwood09a} or the solar panels were included in order to maximize signals and reduce the level of background noise. We used the time-tagged event (TTE) data which provides high temporal (continuous temporal coverage with 2~$\mu$s time tags) and spectral resolution (128 pseudo-logarithmically scaled energy channels). The channels with energy less than 8~keV for NaIs and 245~keV for BGOs, together with the overflow channels, were excluded. As a result, an effective spectral range from 8~keV to 40~MeV was used in the analysis. Moreover, effective area corrections were applied to each pair of NaI and BGO detectors.

\subsection{Time-Resolved Spectral Analysis}
\label{subsect:trsa}

The light curves were binned using a fixed S/N for each burst (but varying across bursts, see last column of Table~\ref{tab:tab1}), in order to avoid artificial binning bias while preserving the general shape of the light curve by avoiding merging peaks and valleys \citep[e.g.][]{Guiriec10a}, resulting in a total of 299 spectra. The binned light curves are shown in Figs.~\ref{fig:AppFig1} and \ref{fig:AppFig2} with time relative to the GBM trigger time $T_0$.

Time-resolved spectroscopy was then performed with the GBM official spectral analysis software RMFIT\footnote{The public version of the RMFIT software is available at http://fermi.gsfc.nasa.gov/ssc/data/analysis/rmfit/} v4.3BA and the GBM response matrices v2.0. In order to account for the change in orientation of the source with respect to the detectors caused by the slew of the spacecraft, RSP2 files containing the detector response matrices (DRM) for every 2 degrees on the sky were used. For each burst a low-order polynomial (order 2 - 4) was fitted to every energy channel according to a user defined background interval before and after the prompt emission phase and interpolated across the emission interval.

\citet{Bhat13a} reported that the typical minimum variability timescales (MVT) for short and long GRBs are 24~ms and 0.25~s respectively. The average temporal resolution of the time bins ($T_\text{bin}$) used in this paper is 2.18~s, which is longer than the MVT. The pulse duration ($T_\text{pulse}$) ranges from seconds to tens of seconds (see Figs.~\ref{fig:AppFig1} and \ref{fig:AppFig2}), which is, of course, by definition shorter than or equal to the burst duration $T_{90}$. So we have the typical values of $\text{MVT}<T_\text{bin}<T_\text{pulse}<T_{90}$.

The variable temporal S/N binning technique can avoid the resulting statistics being dominated by the brightest few bursts. This is because the optimal S/N for each burst is different which lead to similar number of bins for every bursts (see Tables~\ref{tab:taba1} - \ref{tab:taba8}). The fitting results will be given in Sect.~\ref{sect:obs} and discussed in Sect.~\ref{sect:dis}. GRB 100724B will be discussed separately due to its ambiguous parameter distributions. We checked the statistics contributed by individual bursts and found that our conclusions are not affected if any one burst (even for GRB 100724B, see Sect.~\ref{subsect:bandfit}) is removed from the overall sample.

\subsection{Empirical Fitting Models}
\label{subsect:models}

\subsubsection{Band Function (BAND)}

The Band function \citep{Band93a} was fitted to every spectrum:
\begin{equation}
\label{eqn:band}
f_\text{BAND}(E) = A\left\{
\begin{array}{ll}
	\left(\frac{E}{100\text{ keV}}\right)^\alpha \exp\left[-\frac{(\alpha+2)E}{E_\text{p}}\right] \text{ for } E<E_\text{c}, \\
	\left(\frac{E}{100\text{ keV}}\right)^\beta \exp\left(\beta-\alpha\right) \left(\frac{E_\text{c}}{100\text{ keV}}\right)^{\alpha-\beta} \text{ for } E\geq E_\text{c},
\end{array}
\right.
\end{equation}
where
\begin{equation}
\label{eqn:Ec}
E_\text{c}=\left(\frac{\alpha-\beta}{\alpha+2}\right)E_\text{p}.
\end{equation}
In the above equations, $A$ is the normalization factor at 100~keV in units of photons~s$^{-1}$~cm$^{-2}$~keV$^{-1}$, $\alpha$ is the low-energy power-law index, $\beta$ is the high-energy power-law index, and $E_\text{p}$ is the peak energy in units of keV in the observed $\nu F_\nu$ spectrum. The energy $E_\text{c}$ is where the low-energy power-law with an exponential cutoff ends and the pure high-energy power-law starts.

\subsubsection{Synchrotron Model (SYNC)}
\label{subsubsect:syncmodel}

\begin{figure*}
\resizebox{\hsize}{!}
{\includegraphics[width = 18 cm]{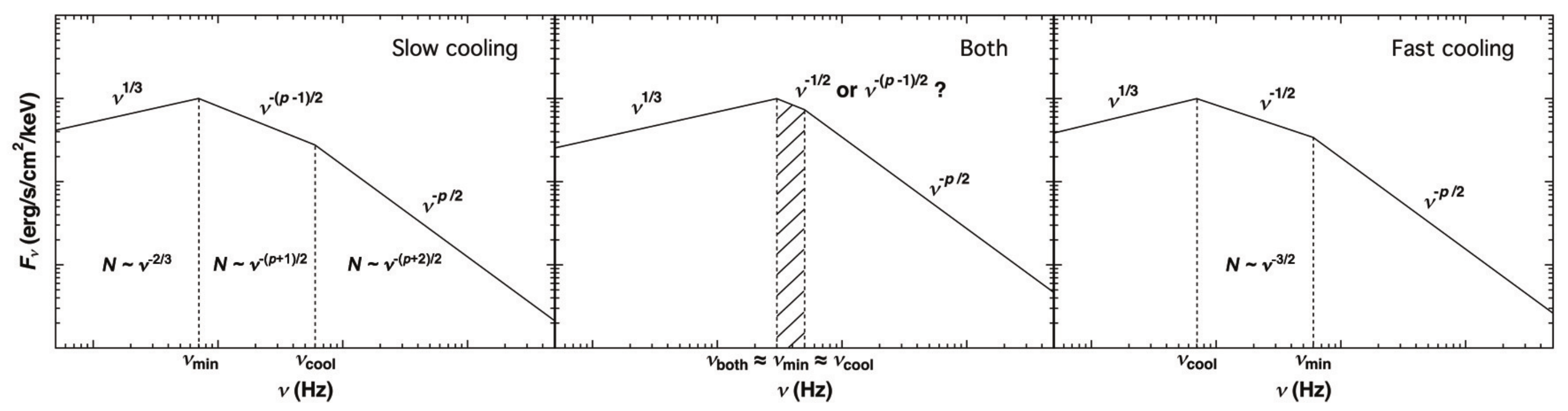}}
\caption{Schematic spectra for the SSM cooling scenarios. The left, middle, and right panels show the "slow", "both", and "fast" cases in the energy flux space, respectively. The shaded region represents the possible location of $\nu_\text{both}$ (i.e. $E_\text{p}$) when fitting the observed spectrum using a model with smoothly jointed power-laws. The photon distribution slopes are also indicated for each different case.}
\label{fig:ssm}
\end{figure*}

The optically thin Synchrotron Shock Model (SSM) predicts two different spectra, "fast-cooling" and "slow-cooling"  \citep[e.g.][]{Sari98a,Preece02a}, depending on the injection and evolution of the relativistic electron population. Both of them consist of a lower and a higher frequency break, fixed by the values of the cooling frequency $\nu_\text{cool}$ and the minimum injection frequency $\nu_\text{min}$ for the relativistic electrons. The electrons in the shock are accelerated to a minimum energy $\gamma_\text{min}$. Assuming a power-law behaviour for the electron energy distribution $N(\gamma_\text{e})\propto\gamma_\text{e}^{-p}$, where $\gamma_\text{e}\geq\gamma_\text{min}$ is the electron energy, the emission spectrum also has a power-law shape. As long as $p > 2$, the distribution is characterized by its lower cut-off at $\gamma_\text{min}$, and the integrated energy of the population does not diverge at high electron energies.

There is a critical energy $\gamma_\text{cool}$ such that electrons with energies above $\gamma_\text{cool}$ emit a significant amount of their energy via synchrotron cooling. The values of $\gamma_\text{cool}$ and $\gamma_\text{min}$ correspond to $\nu_\text{cool}$ and $\nu_\text{min}$ respectively, and the slow-cooling spectrum is given by
\begin{equation}
\label{eqn:slowcool}
F_{\nu\text{,slow}}\propto \left\{
\begin{array}{ll}
	\nu^{1/3} & \text{for } \nu_\text{min} > \nu, \\
	\nu^{-(p-1)/2} & \text{for } \nu_\text{cool} > \nu > \nu_\text{min}, \\
	\nu^{-p/2} & \text{for } \nu > \nu_\text{cool},
\end{array}
\right.
\end{equation}
while the fast-cooling spectrum is given by
\begin{equation}
\label{eqn:fastcool}
F_{\nu\text{,fast}}\propto \left\{
\begin{array}{ll}
	\nu^{1/3} & \text{for } \nu_\text{cool} > \nu, \\
	\nu^{-1/2} & \text{for } \nu_\text{min} > \nu > \nu_\text{cool}, \\
	\nu^{-p/2} & \text{for } \nu > \nu_\text{min}.
\end{array}
\right.
\end{equation}
Subtracting 1 from the spectral indices will give the photon indices (i.e. $\alpha$ and $\beta$) which will be obtained in Sect.~\ref{sect:obs}, leading to a synchrotron "line-of-death" $\alpha = -2/3$ for both scenarios and a second line-of-death $\alpha = -3/2$ \citep{Preece98a} for the fast-cooling scenario. Figure~\ref{fig:ssm} shows the schematic spectra for the slow- and fast-cooling scenario as well as the so-called "both" case where $\nu_\text{cool}/\nu_\text{min}$ (slow-cooling) or $\nu_\text{min}/\nu_\text{cool}$ (fast-cooling) is close to unity. The "both" case can be considered to describe an intermediate case of "moderately fast-cooling".

The synchrotron fitting model that we apply is a modified triple power-law with sharp breaks defined as:
\begin{equation}
\label{eqn:bkn2pow}
f_\text{SYNC}(E) = A\left\{
\begin{array}{ll}
	\left(\frac{E}{100\text{ keV}}\right)^\alpha \text{ for } E < E_\text{b,1}, \\
	\left(\frac{E_\text{b,1}}{100\text{ keV}}\right)^{\alpha-\beta} \left(\frac{E}{100\text{ keV}}\right)^\beta \text{ for } E_\text{b,1} \leq E < E_\text{b,2}, \\
	\left(\frac{E_\text{b,1}}{100\text{ keV}}\right)^{\alpha-\beta} \left(\frac{E_\text{b,2}}{100\text{ keV}}\right)^{\beta-\gamma} \left(\frac{E}{100\text{ keV}}\right)^\gamma \text{ for } E \geq E_\text{b,2},
\end{array}
\right.
\end{equation}
where $A$ is the normalization factor at 100~keV in units of photons~s$^{-1}$~cm$^{-2}$~keV$^{-1}$, $\alpha$, $\beta$, and $\gamma$ are the power-law indices of the three segments (from low to high energies), and $E_\text{b,1}$ and $E_\text{b,2}$ are the two break energies in units of keV. Here we fixed $\alpha=-2/3$ and $\beta-\gamma=1/2$ to create a SYNC-slow model (Eqn.~\ref{eqn:slowcool}). This makes it a four parameter model with freely varying $A$, $E_\text{b,1}$, $E_\text{b,2}$, and $\beta$ (or equivalently, $\gamma$). We also tried to fit the SYNC model with fixed $\alpha=-2/3$ and $\beta=-3/2$ to create a SYNC-fast model (Eqn.~\ref{eqn:fastcool}). This also makes a four parameter model with freely varying $A$, $E_\text{b,1}$, $E_\text{b,2}$, and $\gamma$.

\subsubsection{Blackbody Model (BB)}

We also added a blackbody model to the SYNC fits. It is a Planck function defined as:
\begin{equation}
\label{eqn:bb}
f_\text{BB}(E) = A \left[\frac{(E/1\text{ keV})^2}{\exp(E/kT)-1}\right],
\end{equation}
where $A$ is the normalization factor at 1~keV in units of photons~s$^{-1}$~cm$^{-2}$~keV$^{-1}$ and $kT$ is the temperature of the blackbody in units of keV.

\section{Fitting Results}
\label{sect:obs}

\subsection{BAND Fits}
\label{subsect:bandfit}

\begin{figure*}
\resizebox{\hsize}{!}
{\includegraphics[width = 18 cm]{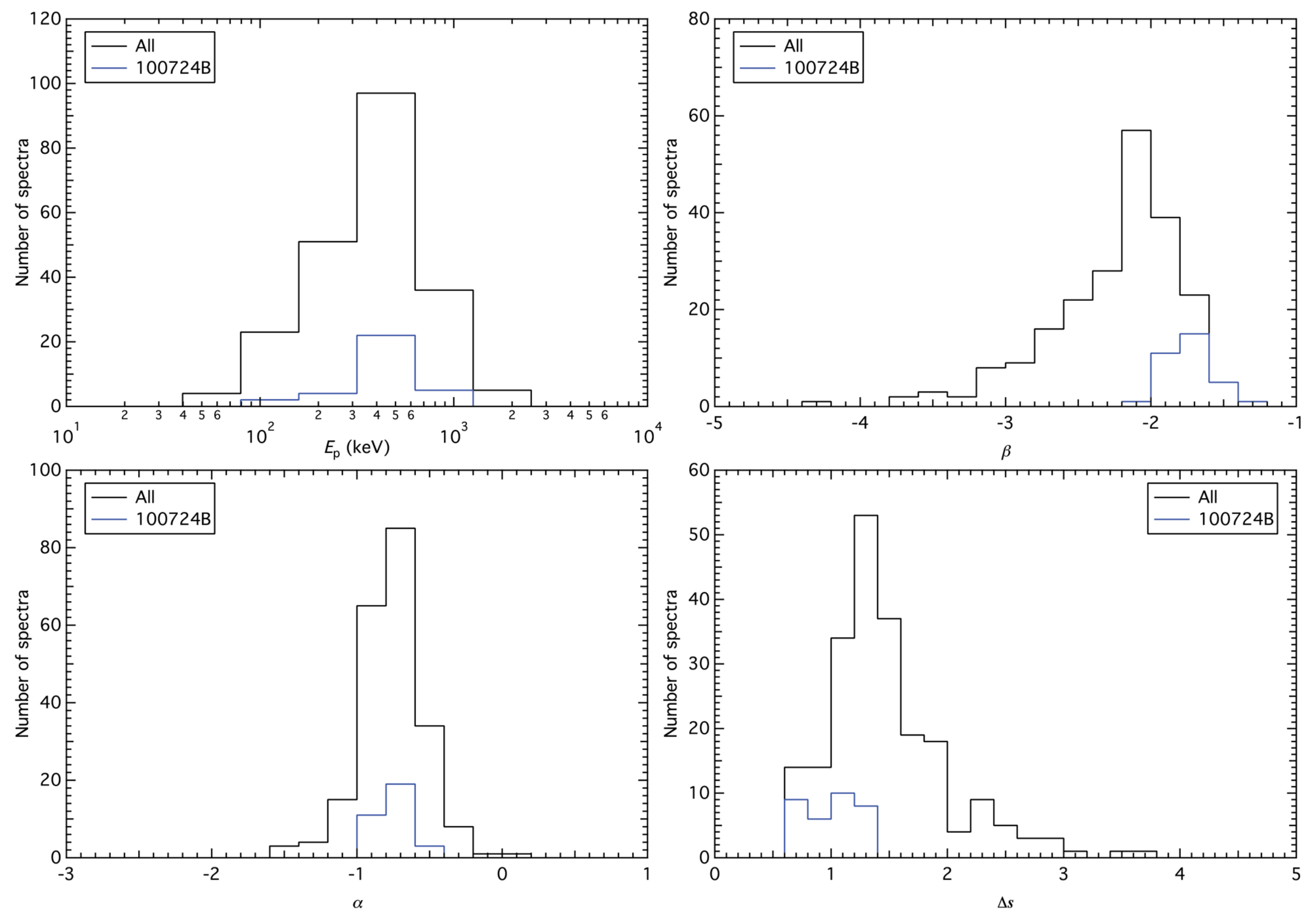}}
\caption{Distributions of the constrained parameters obtained from the BAND model. The upper left panel shows the distributions of the values of $E_\text{p}$. The lower left panel shows the distributions of the values of $\alpha$. The upper right panel shows the distributions of the values of $\beta$. The lower right panel shows the distributions of the values of $\Delta s=\alpha-\beta$. The blue lines show the distributions of GRB 100724B.}
\label{fig:stat1}
\end{figure*}

\begin{figure*}
\resizebox{\hsize}{!}
{\includegraphics[width = 18 cm]{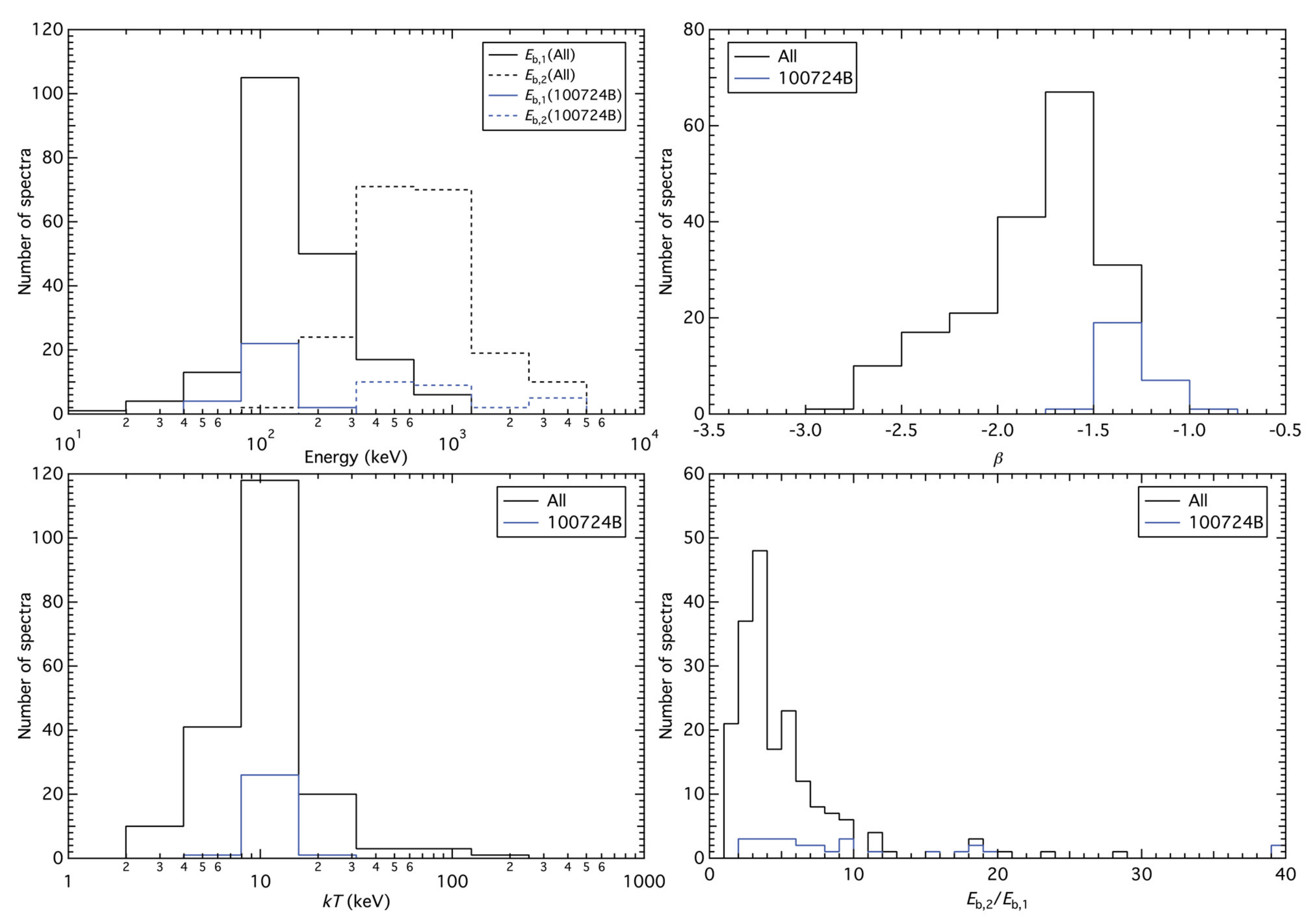}}
\caption{Distributions of the constrained parameters obtained from the SYNC+BB model with slow-cooling constraints (i.e. $\alpha=-2/3$ and $\beta-\gamma=1/2$). The upper left panel shows the break energies $E_\text{b,1}$ and $E_\text{b,2}$. The lower left panel shows the $kT$ distribution. The upper right panel shows the photon indices $\beta$ of the middle power-law segment. The lower right panel shows the ratio between the two breaks, $E_\text{b,2}/E_\text{b,1}$. The blue lines show the distributions of GRB 100724B. Values that are out of the plotting region are accumulated in the boundary bins.}
\label{fig:stat2}
\end{figure*}

The Band function has long been known to provide a good fit to prompt emission spectra \citep{Band93a}, where the typical reduced-$\chi^2 \approx1$ (there is a caveat that the $\chi^2$ statistics may not be suitable for non-Gaussian data) and the Castor C-Statistics values \citep[CSTAT,][]{Cash79a} are low (often a few hundred to a thousand for GBM fits depending on the data quality of individual burst) among the simplest models \citep[e.g.][]{Goldstein12a,Gruber14a,Yu14a}. If, in addition to a low CSTAT value corresponding to a low reduced-$\chi^2$ value ($\approx1$), all parameters in an individual spectral fit have 1-$\sigma$ relative error $\sigma_\text{parameter}/\text{(parameter value)} < 1.0$ (for power-law indices we use absolute error $\sigma_\text{parameter} < 1.0$), we define the fit as a constrained fit. For all these good fits, we verify that the data points are within $\approx99.73$\% confidence level to the model curves. Although we found that in some extreme cases the asymmetric errors of $\beta$ may be unconstrained on the negative side, our selection criteria can filter most of these cases by ensuring the symmetric error (which is the mean of the asymmetric errors) to be well behaved. As a result, 216 of the total 299 spectra ($\approx72\%$) are constrained. Figure~\ref{fig:stat1} shows the distributions of the constrained parameters for the BAND model: the low-energy power-law index $\alpha$, the high-energy power-law index $\beta$, the peak energy in the observed $\nu F_\nu$ spectrum $E_\text{p}$, and the difference between the low- and high-energy power-law indices $\Delta s\equiv\alpha-\beta$.

The distributions of $\alpha$, $\beta$, $E_\text{p}$, and $\Delta s$ are clustered around values of $-0.73^{+0.16}_{-0.21}$, $-2.13^{+0.28}_{-0.56}$, $374.4^{+307.3}_{-187.7}$~keV ($\log_{10}E_\text{p}=2.57^{+0.26}_{-0.30}$), and $1.38^{+0.54}_{-0.31}$, respectively. The asymmetric distribution errors were determined via taking the difference between the median values of the cumulative distribution function (CDF) and the 68\% quantiles. Note that $\alpha=-0.73^{+0.16}_{-0.21}$ shows that the overall sample distribution is consistent with the synchrotron line-of-death (see Sect.~\ref{subsubsect:syncmodel}). About a third of the individual spectra are consistent with the value $\alpha=-2/3$ within 1-$\sigma$. The slope $\beta=-2.13^{+0.28}_{-0.56}$ is consistent with typically observed values. The average errors of $\alpha$ and $\beta$ are $\sigma_\alpha\sim0.1$ and $\sigma_\beta\sim0.2$, respectively. So in Fig.~\ref{fig:stat1} a bin width equals to 0.2 was chosen for displaying the histograms. This implies that the observed dispersions in the power-law index distributions cannot be explained solely by statistical uncertainties. The dispersion is also observed within bursts, indicating that spectral evolution has a non-negligible effect on the parameter distribution. Moreover, it is observed that $\sigma_{E_\text{p}}\sim0.1E_\text{p}$.

The distribution of $E_\text{p}$ peaks at $374.4^{+307.3}_{-187.7}$~keV and are only slightly higher than those found in the GBM time-averaged spectral catalogs \citep{Goldstein13a,Gruber14a} and the BATSE spectral catalogs \citep[e.g.][]{Kaneko06a}. According to Fig.~\ref{fig:stat1}, 91\% of all $E_\text{p}\leq1$~MeV. The remaining 9\% has the highest $E_\text{p}=2.1$~MeV (GRB 130504C, see Table~\ref{tab:taba7}). \citet{Nava11a} presented a time-averaged spectral analysis on 44 short GBM GRBs, and found that the distribution peaks at $E_\text{p}=500^{+260}_{-175}$~keV. This suggests that our long bursts could be as hard as short bursts, which is expected since we selected the bursts with relatively better statistics in the BGO channels. Our bursts lie at the high $E_\text{p}$-long $T_{90}$ end in the long/soft-short/hard classification of GRBs \citep{Kouveliotou93a}. However, it should be noted that the brightest three short GBM GRBs show $E_\text{p}$ as large as 6~MeV \citep{Guiriec10a}. This shows that the $E_\text{p}$ dispersion within long or short bursts can also be huge. In addition, $E_\text{p}$ is observed to be decreasing throughout a burst, with intensity-tracking behaviour during sub-pulses within a single burst (see Sect.~\ref{subsect:htse}).

As shown in Fig.~\ref{fig:stat1}, 50\% of the hard $\beta>-2$ are from GRB 100724B. We will show in Sect.~\ref{subsect:sync} that this burst is consistent with both the slow- and fast-cooling scenario, and that the general conclusion is not affected beacuse removing this burst will only make the distribution peak narrower.

\subsection{SYNC Fits}
\label{subsect:syncfit}

Various studies have shown that a thermal component around a few times 10~keV may generally exist \citep[e.g.][]{Meszaros02a,Ryde05a,Guiriec11a,Axelsson12a,Guiriec13a,Burgess14a,Burgess14b}. In addition, \citet{Burgess14a} showed that a SYNC type model alone cannot be reconciled with the flatness of $\alpha$. We found that in most of our spectra adding a blackbody component can greatly improve the fit. Therefore, all the spectra were fit again to include a blackbody component in the SYNC model. The theoretical implications for the SYNC+BB and BAND model are discussed in Sect.~\ref{sect:dis}.

\begin{figure}
\resizebox{\hsize}{!}
{\includegraphics{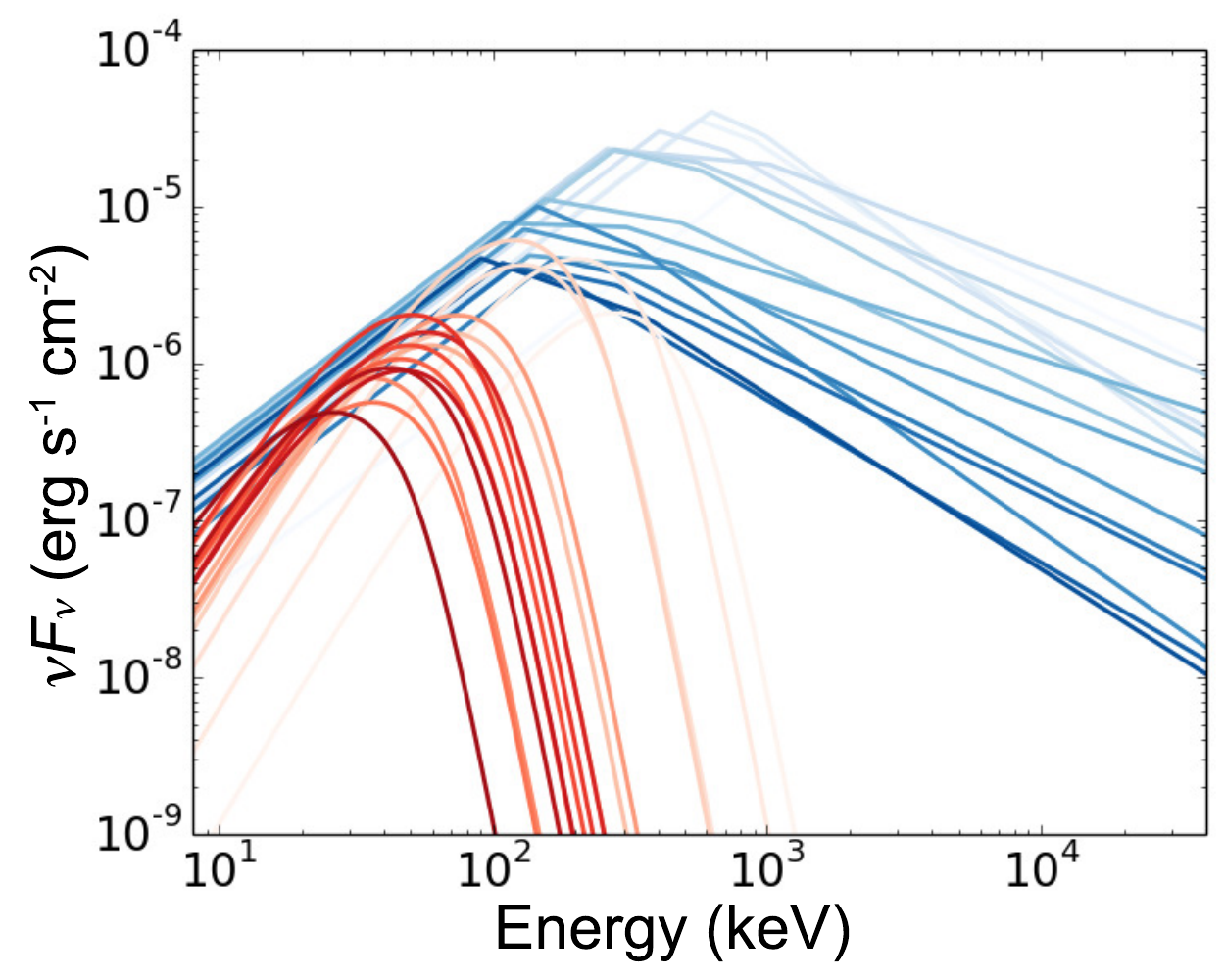}}
\caption{The $\nu F_\nu$ spectral evolution of the SYNC-slow model for GRB 130427A. The evolution of the SYNC component evolves from cyan to blue, while the BB component evolves from yellow to red. No clear correlation is found between the two components.}
\label{fig:evo}
\end{figure}

Two SYNC+BB models (i.e. SYNC-slow+BB and SYNC-fast+BB) were fitted to all spectra using a customized version of RMFIT. We validated that these are good fits to the data by various goodness-of-fit measures: (1) reduced-$\chi^2$ values are close to unity; (2) CSTAT values are comparable to, often lower than, those for the BAND fits \citep[e.g.][]{Gruber14a,Burgess14a}; and (3) quantile-quantile plots for the cumulative observed vs. model count rates lie very close to $x=y$, thus confirming that a SYNC+BB model description is consistent with the data. For reference, the CSTAT values for all spectra are listed together with the degrees of freedom (DOF) and the fitted parameters in Tables~\ref{tab:taba1} - \ref{tab:taba8}. It is found that both the SYNC-slow+BB and -fast+BB models provide constrained fits in more than 65\% of all spectra. We show in Sect.~\ref{sect:dis} that such a test model can provide constraints on various prompt emission mechanism theories. The distributions for the SYNC-slow+BB constrained parameters are plotted in Fig.~\ref{fig:stat2}. The time-resolved spectral evolution for GRB 130427A is shown in Fig.~\ref{fig:evo}. There is no clear correlation found between the fluxes of the SYNC and BB components.

The upper left panel of Fig.~\ref{fig:stat2} shows the distributions of $E_\text{b,1}$ and $E_\text{b,2}$. We found that there are two clear peaks for the breaks around $129.6^{+132.2}_{-32.4}$~keV and $631.4^{+582.6}_{-309.6}$~keV for $E_\text{b,1}$ and $E_\text{b,2}$, respectively. The asymmetric distribution errors were obtained via the same procedure by constructing CDFs as described in Sect.~\ref{subsect:bandfit}. Comparing to the BAND fits, it is observed in most of the cases that $E_\text{b,1}<E_\text{p}\approx E_\text{b,2}$. We found that 100\% of $E_\text{b,1}<1$~MeV and 97\% of $E_\text{b,2}<3$~MeV.

The lower left panel shows the $kT$ distribution. The parameter distribution of $kT=10.4^{+4.9}_{-3.7}$~keV creates a bump at $\sim30$~keV. This $kT$ distribution is consistent with most of the sub-dominant thermal bursts observed \citep[e.g.][]{Ryde05a,Guiriec11a,Axelsson12a,Guiriec13a,Burgess14a,Burgess14b}. When the Planck-to-SYNC flux ratio is high, the Planck function dominates the curvature of the lowest end of the spectrum.

The upper right panel shows the distribution of $\beta$, where $\beta - \gamma = 1/2$. The parameter distribution of $\beta=-1.72^{+0.48}_{-0.25}$ translates to the electron distribution index $p=2.44^{+0.50}_{-0.96}$. A synchrotron spectrum with $p>2$ (i.e. $\beta<-1.5$) requires no upper cut-off in order for the total energy of the electrons to remain finite (Sect.~\ref{subsubsect:syncmodel}). Therefore, the measured high-energy slopes for SYNC model do not require such a cut-off to exist. In addition, this is also consistent with afterglow-deduced distributions of $p\sim2.3$ \citep[e.g.,][]{Curran10a,Ryan14a}. GRB 100724B provided most of the cases where $\beta>-1.5$, which matches the fast-cooling index value.

The lower right panel shows the distribution of the ratio between the two breaks, $E_\text{b,2}/E_\text{b,1}$. It is observed that $E_\text{b,2}$ and $E_\text{b,1}$ have a peak ratio at $3.77^{+4.01}_{-1.53}$, and over 90\% are below 10. If we assume $E_\text{b,1}$ and $E_\text{b,2}$ are related to $E_\text{min}=h\nu_\text{min}$ and $E_\text{cool}=h\nu_\text{cool}$, then a ratio of $E_\text{b,2}/E_\text{b,1}<10$ poses a very tight constraint on the theoretical models (see Sect.~\ref{subsect:sync2}).

The parameter distributions for the SYNC-fast model are nearly identical to those of the SYNC-slow model (which is expected because the value of $\beta=-3/2$ is only 0.1 away from the SYNC-slow $\beta$ distribution peak). The only difference observed is that $\gamma$ extends to much steeper values (from $-1.75$ to $-4.50$ with a peak around $-2.0\text{ - }-2.5$, not a normally distributed population), which reflects the fact that since the power-law segments are no longer connected, $\gamma$ can go much steeper in the time bins that contain mostly upper limits in the high-energy channels.

In brief, the following features are observed in the SYNC fits: (1) over 90\% of $E_\text{b,2}/E_\text{b,1}<10$; (2) a bump/flattening feature at $\sim30$~keV; and (3) a general hard-to-soft evolution for the peak/break energy is observed. We discuss the theoretical implications of these observational results in the next section.

\section{Theoretical Implications}
\label{sect:dis}

\subsection{Hard-to-Soft Evolution and Intensity-Tracking Behaviour}
\label{subsect:htse}

We show the light curves overlaid on the evolutions of $E_\text{p}$, $E_\text{b,1}$, $E_\text{b,2}$, and $kT$ for every burst in Figs.~\ref{fig:AppFig1} and \ref{fig:AppFig2}. Hard-to-soft evolution over the whole bursting period is observed in every burst with in-pulse intensity-tracking behaviour. These two modes of evolutionary trend have been observed in many GRBs \citep[e.g.,][]{Ford95a,Liang96a,Kaneko06a,Preece00a,Guiriec10a,Lu10a,Peng10a,Ghirlanda11a,Burgess14a,Preece14a}. Hard-to-soft evolution is a natural prediction from the SSM \citep{Daigne98a}, in which the relative Lorentz factors of the colliding shells become lower and the spectra become softer. For instance, \citet{Lu12a} reported a time-resolved spectral analysis for 62 \textit{Fermi} bursts (51 long + 11 short) with a detailed study of the $E_\text{p}$ evolution. They found that the two modes for $E_\text{p}$ evolution are present in different pulses and in different bursts. Despite the complexity of the issue, they suggested that the intensity-tracking behaviour could be at least partially attributed to the superposition of hard-to-soft pulses in a highly superimposed light curve. As all bursts in our sample are multi-pulsed (though for GRB 130427A only the first pulse is analyzed, see Sect.~\ref{subsect:data}), this possibility cannot be excluded. We also observed that the $E_\text{p}$ in later pulses never gets as high as in the first pulse, even if a later pulse has a higher peak flux. This suggests that the hard-to-soft evolution dominates over the intensity-tracking behaviour, and that the hard-to-soft evolution is an intrinsic property of GRBs with intensity-tracking behaviour added on top.

\subsection{Synchrotron Emission and the Band Function Fits}
\label{subsect:sync}

\begin{table*}[!htbp]
\caption{Electron distribution index $p$ for different cases.}
\label{tab:tab2}
\centering
\begin{tabular}{cccccccc}
\hline\hline
Case\tablefootmark{a} & $\alpha$ & $p=f(\Delta s)$ & $f(1.2)$ - $f(1.6)$\tablefootmark{b} & $f(1.0)$\tablefootmark{c} & $p=g(\beta)$ & $g(-2.0)$ - $g(-2.4)$\tablefootmark{b} & $g(-1.7)$\tablefootmark{c} \\
(1) & (2) & (3) & (4) & (5) & (6) & (7) & (8) \\
\hline
Fast, high... & $-3/2$ & $2\Delta s + 1$ & 3.4 - 4.2 & 3.0 & $-2(\beta+1)$ & 2.0 - 2.8 & 1.4 \\
Slow, low... & $-2/3$ & $2(\Delta s + 1/6)$ & 2.73 - 3.53  & 2.33 & $-2\beta-1$ & 3.0 - 3.8 & 2.4 \\
Both........... & $-2/3$ & $2(\Delta s - 1/3)$ & 1.73 - 2.53 & 1.33  & $-2(\beta+1)$ & 2.0 - 2.8 & 1.4 \\
\hline
\end{tabular}
\tablefoot{
\tablefoottext{a}{\citet{Preece02a}, Eqns. (9), (10), and (12).}
\tablefoottext{b}{Calculated from the ranges of peak and average values of $\Delta s$ and $\beta$ distributions for all eight bursts, given that $1.2<(\Delta s)_\text{peak}<1.4$, $1.4<(\Delta s)_\text{average}<1.6$, $-2.2<\beta_\text{peak}<-2.0$, and $-2.4<\beta_\text{average}<-2.2$.}
\tablefoottext{c}{Calculated from the average values of $\Delta s$ and $\beta$ distributions for GRB 100724B only.}
}
\end{table*}

The values of $\Delta s$ and $\beta$ obtained from the BAND fits can be used to compute the electron distribution power-law index $p$ and to distinguish among different cooling scenarios \citep{Preece02a}. \citet{Preece02a} performed time-resolved spectroscopy on 156 BATSE GRBs and found that the results are consistent with the "slow, low", "both", or "fast, high" cases (with "low" and "high" referring to just the lower or higher spectral break respectively).

The relative rate of electron cooling against energy injection into the electron population marks the difference between slow- and fast-cooling. To obtain the synchrotron cooling and energy injection timescale requires knowledge of the physical parameters of the ejecta, e.g. magnetic field strength and electron Lorentz factor, as well as precise modelling of the energy output from the central engine. This makes accurate measurements of these timescales difficult. In the internal shock model, the relative Lorentz factors between colliding shells are only mildly relativistic \citep{Daigne98a}, and the synchrotron cooling timescale of the relativistic electrons in the ejecta frame is
\begin{equation}
t_\text{syn} = 6 \left(\frac{\Gamma_\text{e}}{100}\right)^{-1}\left(\frac{B}{1000\text{ G}}\right)^{-2} \text{ s},
\end{equation}
where $\Gamma_\text{e}$ is the Lorentz factor of the electrons relative to the ejecta and $B$ is the magnetic field in the shock. One could compare, for instance, $t_\text{syn}$ with the MVT observed in the light curve (as experienced in the ejecta frame), which is then taken to represent the rate of energy injection into the synchrotron electron population. However, the inferred values of the physical parameters, such as magnetic field strength (see the discussion below), vary in a wide range among bursts and sub-pulses within a single bursts. Taken together with the uncertainty in the spatial and temporal profile of the particle acceleration sites (e.g. extended turbulent regions vs. shock acceleration, or intermittent vs. continuous injection), it becomes hard to predict a clear preference for a given cooling regime due to the difficulty of unambiguously interpreting the observable time scales. We show in the following that a mix of both the slow- and fast-cooling is implied by the GBM data.

Table~\ref{tab:tab2} shows the values of $p$ obtained from the $\Delta s$ and $\beta$ distributions \citep[see Eqns. 8 - 12 in][]{Preece02a}. Column 1 shows the three cases where $p$ depends on both $\Delta s$ and $\beta$. Column 2 shows the respective value of $\alpha$ in each case. Columns 3 and 6 show the formulae for $p$ as a function of $\Delta s$ and $\beta$ respectively. Columns 4 and 7 give the ranges of possible values of $p$ calculated from the distributions of $\Delta s$ and $\beta$ for all eight bursts, and Cols. 5 and 8 give the same for GRB 100724B alone.

It can be seen that the values of $p$ in Col. 4 are inconsistent with the "fast, high" case in Col. 7. The "fast, low" case predicts $\Delta s=5/6$ which is clearly rejected as shown in Fig.~\ref{fig:stat1}. The distribution of fast-cooling $\gamma_\text{SYNC}$ as mentioned in Sect.~\ref{subsect:syncfit} indicates that the electron distribution index above $\gamma_\text{min}$ can take any value from $p=1.5\text{ - }6.0$. Theories of electron shock-acceleration typically predict $p$ values between 2 and 3, which makes these very steep values for $p$ suggestive of the presence of a cut-off or deviation from a power-law slope in the accelerated particle distribution, rather than a single very steep slope. A steep electron distribution index can also occur when the shock normal is at an angle to the magnetic field, allowing electrons to escape the acceleration region early \citep{Ellison04a,Baring06a,Summerlin12a,Burgess14a}. The "slow, high" case, which refers to the higher energy break in the left panel of Fig.~\ref{fig:ssm}, predicts $\Delta s = 1/2$ and is clearly rejected as shown in Fig.~\ref{fig:stat1}. The average values of $\Delta s$ and $\beta$ for GRB 100724B are 1.0 and $-$1.7 respectively, which are also consistent with the "slow, low" and "both" cases (Cols. 5 and 8), at the same time consistent with the "fast, low" case which predicts $\Delta s=5/6$.

On the other hand, the BATSE $\beta$ and $\Delta s$ distributions suggested that the "slow, low", "fast, low", and "both" cases are all viable processes \citep[see, e.g., Fig.~2 of][]{Preece02a,Kaneko06a}. \citet{Gruber14a} also showed similar conclusions in the GBM time-averaged spectra. \citet{Burgess14a} performed a Bayesian time-resolved spectral analysis using physical synchrotron and thermal models instead of the Band function to several GBM GRBs and found that the slow-cooling scenario is a better explanation to the observed data, and their results suggest continuous energy injection is important. \citet{Uhm14a} predicted that using a decaying magnetic field as a function of radius, with a decay index $b$, it is possible for most GRBs to cool via the fast-cooling scenario with $\alpha\sim-1.0$. They predicted that the asymptotic value of the low-energy electron distribution should be $p=(6b-4)/(6b-1)$ instead of $p=2$ for a constant magnetic field \citep[e.g.][]{Preece02a}, and the spectral index $s=(-p+1)/2=3/(12b-2)=\alpha+1$. We found that in more than 77\% of the constrained fits $b$ has values between 0.6 and 2.6. There is no clear evolutionary trend of $b$. The variability of $b$ within bursts is difficult to reconcile with a large scale power-law dependence on radius of the magnetic field. However, this can still be the case, but just not as clearly manifested in the data as predicted by \citet{Uhm14a}.

In brief, our results are consistent with slow-cooling with the low-frequency break seen (or in the "both" case, undistinguished between slow- and fast-cooling). In the case of GRB 100724B, fast-cooling is also consistent with the low-frequency break seen. This implies that the second line-of-death, $\alpha=-3/2$, could also be avoided.

\subsection{Synchrotron Models Fits}
\label{subsect:sync2}

The SYNC-slow model is basically a three-segment broken power-law, with the middle- and high-energy segment connected (i.e. $\beta_\text{SYNC}-\gamma_\text{SYNC}=1/2$). It is essentially an extended version of the BAND model, in which the curvature of BAND is replaced by two breaks and the power-law segment in between. This implies that when we are comparing the results from BAND and SYNC fits, it should be kept in mind that either $\beta_\text{SYNC}$ or $\gamma_\text{SYNC}$ could be picking up $\beta_\text{BAND}$. This is discussed later in the current subsection. It should also be noted that a sharply joined broken power-law is intrinsically non-physical. The actual spectrum should always be smooth, so sharp power-law fits run the risk of covering a single smooth transition with multiple sharp breaks. However, a smoothly joined triple power-law would contain too many parameters and to fit such a complicated empirical model is statistically unsound. The constrained power-law indices in our SYNC-slow and -fast models mitigate the issue by assuming a synchrotron origin of the observed spectrum a priori, thereby limiting the possible shapes of fitted spectra.

Theoretically, the SYNC model has excluded the synchrotron emission from Maxwellian electrons. This is because we wanted to have the synchrotron emission occurring at the right frequencies (i.e. $\gamma$-rays), which requires the energy per emitting electron to be higher than that obtained by simply averaging \citep[as demonstrated by][]{Daigne98a}. This implies a small subset of electrons at very high energies, far away from the thermal pool from which they were drawn. Alternatively, if the Maxwellian electron distribution peak and the minimum injection Lorentz factor (i.e. $\gamma_\text{min}$) remain close, the effect of adding the Maxwellian electrons will be a smoothening of the synchrotron function that we could not model by our BAND or SYNC models. Moreover, \citet{Burgess11a} has shown that the Maxwellian electron population is sub-dominant. Thus in order to avoid further complication of the fitting model, we assume the synchrotron emission is just from the population of shock-accelerated electrons (see Sect.~\ref{subsect:syncfit}). However, one should note also that the Maxwellian does not just have to exist as left over thermal pool from the thermal parts of the jet. It can also be created in the shock region due to thermalization of electrons crossing the shock \citep[see, e.g.,][]{Spitkovsky08a}.

According to the SYNC-slow fitting results, there are two cases to consider: (1) the $\gamma_\text{SYNC}$ is the high-energy segment in the slow cooling scenario, i.e. $\nu_\text{min}$ and $\nu_\text{cool}$ are the predicted break values; or (2) $\gamma_\text{SYNC}$ is the middle-energy segment in the slow-cooling scenario, in this case the triple power-law is just mimicking the slowly varying BAND model. If (1) is true, then we can take $\gamma_\text{SYNC} = -2.5\text{ - }-2.0$, and we will have $p=2.0\text{ - }3.0$. Looking at Table~\ref{tab:tab2}, it can be seen that the SYNC-slow model is consistent with the "both" case; if (2) is true, then instead of comparing to $\gamma_\text{SYNC}$, we should compare with $\beta_\text{SYNC}$ in Eqn.~\ref{eqn:slowcool}, and we will have $p=3.0\text{ - }4.0$. Looking at Table~\ref{tab:tab2}, it can be seen that the SYNC-slow model is also consistent with the "slow, low" case.

\begin{figure}
\resizebox{\hsize}{!}
{\includegraphics{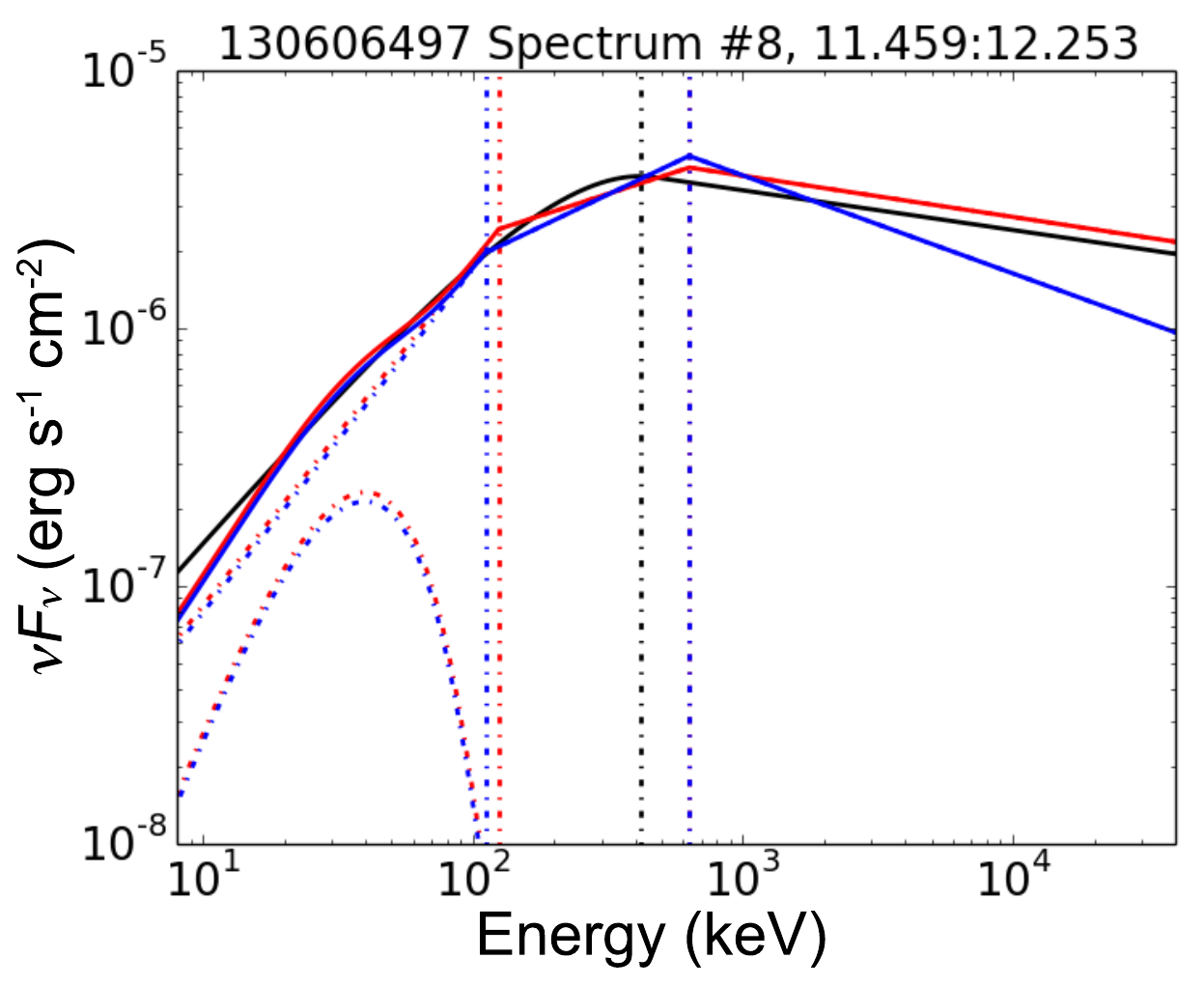}}
\caption{A selected spectrum from GRB 130606B plotted in $\nu F_\nu$ space. The black, red, and blue solid curve show the fitted spectrum for the BAND, SYNC-slow+BB, and SYNC-fast+BB model, respectively, while the dash-dotted curves show individual SYNC or BB component. The vertical dash-dotted black, red, and blue line show the $E_\text{p}$ and break energies for the BAND, SYNC-slow+BB, and SYNC-fast+BB model, respectively.}
\label{fig:spec}
\end{figure}

\citet{Burgess14a} used a physical non-thermal plus thermal synchrotron kernel to fit a few GBM GRBs and found that slow-cooling is physically possible. Since the typically observed value of $\alpha\sim -1.0$, the fast-cooling model has been disfavoured as it predicts $\alpha$ should be as steep as $-3/2$ below $\nu_\text{min}$ \citep{Sari98a}. The presence of a blackbody contribution to the lower part of the spectrum would render it even more difficult to reconcile the $\alpha$ slope with the "fast, high" case. On the one hand, our fit results for the SYNC-slow model yield $p$ values closer to the expected range between 2 and 3. On the other hand, a SYNC-fast model, implying that most of the energy of the electrons is radiated away, has the advantage of allowing for a lower efficiency. The total energy in $\gamma$-rays is typically comparable to the inferred kinetic energy of the ejecta. Therefore, if the efficiency in converting accelerated electron energies to radiation is low, the efficiency in extracting energy from the ejecta to the non-thermal electron population has to become extremely high in order to compensate \citep[see e.g.][for detailed discussions]{Nousek06a,Granot06a}. The fact that in the SYNC fits, both spectral breaks consistently occur fairly close to one another, does alleviate the issue, in that it provides essentially a "moderately fast-cooling" scenario, regardless of the precise order of the breaks. This, however, begs the question how to understand the universal break ratio between $\nu_\text{min}$ and $\nu_\text{cool}$ inferred from our sample, as the positions of these breaks are not theoretically expected to be related.

The fast-cooling model with a decaying magnetic field \citep{Uhm14a} predicts a Band function spectral shape with $b\sim1.0\text{ - }1.5$ (see their Fig.~4), in which the curved Band shape is a sum-up effect for the emissions of electrons at different times. A decay index $b \lesssim 2.6$ (see Sect.~\ref{subsect:sync}) implies stronger magnetic dissipation and the electrons at later time could be cooled via slow-cooling, thus the positions of $\nu_\text{cool}$ and $\nu_\text{min}$ could reverse and move closer to each other, so that the "both" case is possible. \citet{Uhm14a} showed that this is possible in a timescale $\sim1.0$~s, consistent with the typical $T_\text{bin}$ used in this paper (see Sect.~\ref{subsect:trsa}). We found that the BAND and SYNC model have extremely similar shapes (Fig.~\ref{fig:spec}), consistent with this interpretation and thus providing further support for the "both" case.

\subsection{Thermal Origin of Prompt Emission}
\label{subsect:thermal}

Recently, \citet{Beloborodov13a} suggested that the evolution of $E_\text{p}$ could be a manifestation of thermal emission. As shown in Fig.~\ref{fig:stat1}, more than 90\% of $E_\text{p}$ values remain below 1~MeV. The observed clustering of $E_\text{p}\sim$ few hundred keV, instead of a wide distribution, is hard to explain in the SSM. The observed spectral width in the $\nu F_\nu$ space, is $\log (E_1/E_2) \approx 1.0 \text{ - } 1.5$ decades in photon energy \citep{Beloborodov13a}, where $E_2 - E_1$ is the width at half-maximum. This is narrower than a synchrotron model would predict \citep{Daigne11a}.

Early photospheric models assumed a freely expanding radiation-dominated outflow with no baryonic loading or magnetic field \citep{Goodman86a,Paczynski86a}. This predicts a sharply defined peak with a Planck spectrum \citep{Beloborodov11a}, which is in contradiction to the observed non-thermal spectra in most GRBs. Detailed radiation transfer simulations have shown that a thermal origin of the Band function is possible \citep{Peer06b,Giannios08a,Beloborodov10a,Vurm11a}, and \citet{Beloborodov13a} computed that the maximum $E_\text{p}$ of a spectrum from thermal plasma is given by $30\Gamma_\text{P}$~keV under high radiation efficiency, where $\Gamma_\text{P}$ is the Lorentz factor of the Planckian photospheric shell. With the typical values of the Lorentz factor of GRBs to be $\sim100$, $E_\text{p,max}\sim3$~MeV in the rest frame. This value is consistent with most of our observed $E_\text{p}\lesssim500$~keV, but only when assuming a redshift $z\lesssim0.83$. \citet{Deng14a} also found that $\alpha\sim-1.0$ could be achieved if the radiating photosphere has a constant or increasing luminosity. However, they stated that it is difficult to reproduce the observed hard-to-soft evolution under natural conditions.

\section{Conclusions}
\label{sect:conc}

We performed time-resolved spectroscopy for eight energetic, long GRBs observed by \textit{Fermi} GBM during the first five years of its mission. We obtained well constrained BAND spectral parameters and studied their theoretical implications. We showed that even in the bursts with good high-energy statistics above 900~keV, most observed properties can be explained using the Synchrotron Shock Model. We further tested the observed spectra with a synchrotron plus blackbody model using slow- and fast-cooling parametric constraints, and found that the "both" case is consistent with the data, which requires a narrow distribution of the break ratio $E_\text{b,2}/E_\text{b,1}<10$ with a peak at $3.77^{+4.01}_{-1.53}$. The population of $p$ is found to be 2 - 3, in accordance with the expected range. The picture of a "moderately fast-cooling" scenario can also explain the narrow distribution of the break ratio and relax the efficiency issue for the slow-cooling scenario.

Recently, \citet{Frontera13a} reported the result of the time-resolved spectral analysis of four GRBs observed by BATSE and \textit{BeppoSAX}. They found that a specially devised empirical Comptonized model is the best fit model to most of their time-resolved spectra. They also found that using a simple power-law plus blackbody model (PL$+$BB) does not give fitting results better than the conventional BAND function. This is consistent with the results from the time-resolved GBM GRB catalog \citep{Yu14a} that most of the time-resolved spectra are best fitted by a Comptonized model, and only very few spectra are best fitted by PL$+$BB although they are generally not bad fits. We showed in this paper that the spectral shape $\gtrsim1$~MeV could be harder than a Comptonized model or simple power-law.

Our results confirmed that while most properties of energetic GRBs can be explained in the conventional theoretical models, the radiative process in GRB prompt emission is complicated and cannot be fully explained by a single distribution of electrons (e.g. due to anisotropic distribution of electron energies or continuous acceleration or photospheric emission). The possibility of a decaying magnetic field which modifies the fast-cooling spectrum is also explored, yielding a magnetic field decay index $0.6<b<2.6$ for 77\% of the constrained fits. 'However, it is difficult to reconcile the variability of $b$ within bursts with a mechanism where the spectra are shaped by a single large scale decaying magnetic field. Nevertheless, such a field might still exist, but with its impact obscured by more local conditions in the flow.

\begin{acknowledgements}

The authors want to thank Fr{\'e}d{\'e}ric Daigne, Alexander van der Horst, Re'em Sari, Bing Zhang, and the anonymous referee for insightful suggestions. HFY and JG acknowledge support by the DFG cluster of excellence “Origin and Structure of the Universe” (www.universe-cluster.de). The GBM project is supported by the German Bundesministeriums f{\"u}r Wirtschaft und Technologie (BMWi) via the Deutsches Zentrum f{\"u}r Luft und Raumfahrt (DLR) under the contract numbers 50 QV 0301 and 50 OG 0502.

\end{acknowledgements}

\bibliographystyle{aa} 
\bibliography{mybib} 

\begin{appendix}

\section{Time-Resolved Fitting Results}
\label{app:trpara}

\begin{table*}[!htbp]
\tiny
\caption{BAND Parameters for GRB 090902B. The times $t_\text{start}$ and $t_\text{stop}$ are relative to the GBM trigger time $T_0$.}
\label{tab:taba1}
\begin{tabular}{ccccccc}
\hline\hline
$t_\text{start}$ & $t_\text{stop}$ & $A$ & $E_\text{p}$ & $\alpha$ & $\beta$ & CSTAT/DOF \\
(s) & (s) & (ph~s$^{-1}$~cm$^{-2}$~keV$^{-1}$) & (keV) &  &  & \\
\hline
0.000 & 2.325 & 0.0681$\pm$0.0043 & 521.5$\pm$24.1 & $-$0.307$\pm$0.039 & $-$4.168$\pm$1.980 & 539.10/478 \\
2.325 & 4.207 & 0.0745$\pm$0.0043 & 589.1$\pm$27.2 & $-$0.220$\pm$0.041 & $-$3.158$\pm$0.365 & 543.25/478 \\
4.207 & 5.987 & 0.0780$\pm$0.0041 & 620.5$\pm$24.7 & $-$0.197$\pm$0.038 & $-$4.367$\pm$1.810 & 547.67/478 \\
5.987 & 7.173 & 0.0885$\pm$0.0038 & 959.9$\pm$41.2 & $-$0.366$\pm$0.030 & $-$4.600$\pm$1.470 & 662.97/478 \\
7.173 & 7.892 & 0.1352$\pm$0.0054 & 1421.0$\pm$79.2 & $-$0.782$\pm$0.019 & $-$5.175$\pm$3.080 & 799.87/478 \\
7.892 & 8.340 & 0.2221$\pm$0.0093 & 1631.0$\pm$128.0 & $-$1.099$\pm$0.016 & $<-$15.53 & 826.49/478 \\
8.340 & 8.738 & 0.2148$\pm$0.0100 & 1811.0$\pm$198.0 & $-$1.153$\pm$0.016 & $-$3.457$\pm$0.674 & 938.64/478 \\
8.738 & 9.176 & 0.2253$\pm$0.0095 & 1801.0$\pm$161.0 & $-$1.169$\pm$0.015 & $<-$5.522 & 974.42/478 \\
9.176 & 9.554 & 0.2497$\pm$0.0112 & 1737.0$\pm$181.0 & $-$1.279$\pm$0.015 & $<-$5.696 & 764.26/478 \\
9.554 & 9.878 & 0.3091$\pm$0.0163 & 1082.0$\pm$114.0 & $-$1.201$\pm$0.018 & $-$3.909$\pm$1.280 & 867.81/478 \\
9.878 & 10.262 & 0.2456$\pm$0.0123 & 1171.0$\pm$106.0 & $-$1.123$\pm$0.018 & $<-$12.89 & 843.17/478 \\
10.262 & 10.730 & 0.2260$\pm$0.0095 & 1588.0$\pm$133.0 & $-$1.147$\pm$0.016 & $<-$8.822 & 1000.1/478 \\
10.730 & 11.116 & 0.2432$\pm$0.0117 & 1324.0$\pm$123.0 & $-$1.108$\pm$0.018 & $-$4.184$\pm$1.630 & 779.39/478 \\
11.116 & 11.633 & 0.1634$\pm$0.0083 & 1272.0$\pm$119.0 & $-$1.110$\pm$0.018 & $<-$12.32 & 805.70/478 \\
11.633 & 12.270 & 0.1398$\pm$0.0077 & 1226.0$\pm$137.0 & $-$1.140$\pm$0.020 & $-$4.258$\pm$2.460 & 712.95/478 \\
12.270 & 12.982 & 0.2276$\pm$0.0359 & 137.4$\pm$12.3 & $-$0.969$\pm$0.062 & $-$2.214$\pm$0.106 & 557.08/478 \\
12.982 & 13.337 & 0.6174$\pm$0.0740 & 212.2$\pm$11.7 & $-$0.495$\pm$0.051 & $-$2.451$\pm$0.110 & 579.38/478 \\
13.337 & 13.799 & 0.2969$\pm$0.0318 & 297.8$\pm$20.7 & $-$0.808$\pm$0.039 & $-$2.751$\pm$0.257 & 475.17/478 \\
13.799 & 14.247 & 0.2260$\pm$0.0131 & 668.4$\pm$38.6 & $-$0.776$\pm$0.025 & $<-$11.92 & 536.24/478 \\
14.247 & 14.773 & 0.1927$\pm$0.0117 & 676.2$\pm$42.1 & $-$0.810$\pm$0.025 & $<-$9.535 & 526.75/478 \\
14.773 & 15.186 & 0.2159$\pm$0.0117 & 746.0$\pm$39.0 & $-$0.665$\pm$0.025 & $<-$11.92 & 552.70/478 \\
15.186 & 15.682 & 0.1805$\pm$0.0083 & 999.9$\pm$53.3 & $-$0.628$\pm$0.024 & $-$4.202$\pm$0.990 & 622.98/478 \\
15.682 & 16.280 & 0.1591$\pm$0.0114 & 603.1$\pm$46.8 & $-$0.820$\pm$0.029 & $-$2.772$\pm$0.268 & 566.78/478 \\
16.280 & 16.753 & 0.2031$\pm$0.0118 & 666.4$\pm$37.1 & $-$0.683$\pm$0.027 & $-$4.580$\pm$2.870 & 586.40/478 \\
16.753 & 17.418 & 0.2017$\pm$0.0190 & 366.7$\pm$26.3 & $-$0.705$\pm$0.039 & $-$2.562$\pm$0.185 & 557.65/478 \\
17.418 & 18.232 & 0.1534$\pm$0.0144 & 382.1$\pm$29.3 & $-$0.822$\pm$0.037 & $-$2.699$\pm$0.284 & 467.21/478 \\
18.232 & 18.977 & 0.1515$\pm$0.0136 & 427.2$\pm$34.5 & $-$0.907$\pm$0.033 & $-$2.916$\pm$0.485 & 484.31/478 \\
18.977 & 19.575 & 0.2776$\pm$0.0310 & 278.2$\pm$19.1 & $-$0.570$\pm$0.049 & $-$2.285$\pm$0.096 & 491.66/478 \\
19.575 & 19.995 & 0.3303$\pm$0.0322 & 343.4$\pm$22.5 & $-$0.606$\pm$0.041 & $-$2.563$\pm$0.156 & 509.15/478 \\
19.995 & 20.571 & 0.1728$\pm$0.0161 & 419.3$\pm$36.4 & $-$0.869$\pm$0.036 & $-$2.450$\pm$0.180 & 509.15/478 \\
20.571 & 21.148 & 0.3419$\pm$0.0417 & 217.2$\pm$13.7 & $-$0.645$\pm$0.050 & $-$2.397$\pm$0.119 & 548.28/478 \\
21.148 & 21.843 & 0.3242$\pm$0.0417 & 197.3$\pm$11.6 & $-$0.524$\pm$0.055 & $-$2.395$\pm$0.121 & 564.19/478 \\
21.843 & 23.098 & 0.0832$\pm$0.0103 & 342.9$\pm$45.0 & $-$1.146$\pm$0.040 & $-$2.283$\pm$0.192 & 565.37/478 \\
23.098 & 33.792 & 0.0184$\pm$0.0055 & 134.3$\pm$19.5 & $-$1.427$\pm$0.061 & $-$2.727$\pm$0.780 & 627.70/478 \\
\hline
\end{tabular}
\end{table*}

\clearpage

\begin{table*}[!htbp]
\tiny
\caption{BAND Parameters for GRB 100724B. The times $t_\text{start}$ and $t_\text{stop}$ are relative to the GBM trigger time $T_0$.}
\label{tab:taba2}
\begin{tabular}{ccccccc}
\hline\hline
$t_\text{start}$ & $t_\text{stop}$ & $A$ & $E_\text{p}$ & $\alpha$ & $\beta$ & CSTAT/DOF \\
(s) & (s) & (ph~s$^{-1}$~cm$^{-2}$~keV$^{-1}$) & (keV) &  &  & \\
\hline
-7.168 & 10.075 & 0.0079$\pm$0.0005 & 961.0$\pm$129.0 & $-$0.878$\pm$0.032 & $-$1.581$\pm$0.032 & 22341./478 \\
10.075 & 12.503 & 0.0358$\pm$0.0024 & 570.7$\pm$56.3 & $-$0.698$\pm$0.043 & $-$1.721$\pm$0.035 & 2558.6/478 \\
12.503 & 15.016 & 0.0292$\pm$0.0019 & 662.1$\pm$65.9 & $-$0.727$\pm$0.040 & $-$1.722$\pm$0.037 & 2459.2/478 \\
15.016 & 17.022 & 0.0306$\pm$0.0020 & 1222.0$\pm$161.0 & $-$0.877$\pm$0.030 & $-$1.790$\pm$0.051 & 1297.9/478 \\
17.022 & 19.118 & 0.0301$\pm$0.0020 & 968.7$\pm$114.0 & $-$0.793$\pm$0.034 & $-$1.744$\pm$0.043 & 1608.2/478 \\
19.118 & 21.649 & 0.0249$\pm$0.0015 & 940.3$\pm$115.0 & $-$0.882$\pm$0.033 & $-$1.648$\pm$0.035 & 3636.6/478 \\
21.649 & 24.481 & 0.0262$\pm$0.0021 & 580.1$\pm$83.4 & $-$0.847$\pm$0.044 & $-$1.580$\pm$0.029 & 4363.6/478 \\
24.481 & 29.293 & 0.0208$\pm$0.0017 & 473.6$\pm$80.8 & $-$0.862$\pm$0.051 & $-$1.494$\pm$0.022 & 10681./478 \\
29.293 & 38.468 & 0.0150$\pm$0.0017 & 282.6$\pm$49.7 & $-$0.888$\pm$0.065 & $-$1.511$\pm$0.015 & 53566./478 \\
38.468 & 41.089 & 0.0445$\pm$0.0052 & 286.5$\pm$38.3 & $-$0.688$\pm$0.067 & $-$1.595$\pm$0.024 & 6721.3/478 \\
41.089 & 45.942 & 0.0249$\pm$0.0025 & 298.9$\pm$36.6 & $-$0.787$\pm$0.058 & $-$1.628$\pm$0.020 & 20827./478 \\
45.942 & 50.235 & 0.0184$\pm$0.0018 & 375.9$\pm$52.8 & $-$0.890$\pm$0.052 & $-$1.636$\pm$0.022 & 20121./478 \\
50.235 & 55.075 & 0.0179$\pm$0.0017 & 352.5$\pm$44.9 & $-$0.877$\pm$0.051 & $-$1.646$\pm$0.019 & 20911./478 \\
55.075 & 57.006 & 0.0397$\pm$0.0030 & 525.8$\pm$43.7 & $-$0.780$\pm$0.039 & $-$1.872$\pm$0.030 & 2355.2/478 \\
57.006 & 58.742 & 0.0420$\pm$0.0034 & 499.2$\pm$43.7 & $-$0.754$\pm$0.041 & $-$1.888$\pm$0.028 & 3499.4/478 \\
58.742 & 59.914 & 0.0677$\pm$0.0053 & 436.5$\pm$37.5 & $-$0.717$\pm$0.044 & $-$1.934$\pm$0.043 & 2615.2/478 \\
59.914 & 60.965 & 0.0751$\pm$0.0072 & 364.7$\pm$34.8 & $-$0.661$\pm$0.052 & $-$1.936$\pm$0.049 & 2171.3/478 \\
60.965 & 62.298 & 0.0613$\pm$0.0040 & 457.8$\pm$29.2 & $-$0.687$\pm$0.040 & $-$1.978$\pm$0.021 & 2335.1/478 \\
62.298 & 63.162 & 0.0950$\pm$0.0095 & 454.5$\pm$47.1 & $-$0.657$\pm$0.048 & $-$1.903$\pm$0.050 & 829.13/478 \\
63.162 & 64.192 & 0.0869$\pm$0.0099 & 368.5$\pm$38.1 & $-$0.587$\pm$0.056 & $-$1.796$\pm$0.031 & 1239.1/478 \\
64.192 & 65.463 & 0.0752$\pm$0.0075 & 323.4$\pm$30.0 & $-$0.615$\pm$0.056 & $-$1.797$\pm$0.027 & 1907.0/478 \\
65.463 & 66.735 & 0.0689$\pm$0.0056 & 434.0$\pm$39.4 & $-$0.706$\pm$0.045 & $-$1.906$\pm$0.042 & 1695.8/478 \\
66.735 & 67.858 & 0.0807$\pm$0.0082 & 389.2$\pm$38.6 & $-$0.661$\pm$0.051 & $-$1.844$\pm$0.032 & 1289.5/478 \\
67.858 & 69.182 & 0.0672$\pm$0.0064 & 349.3$\pm$32.5 & $-$0.649$\pm$0.053 & $-$1.805$\pm$0.025 & 1703.0/478 \\
69.182 & 70.591 & 0.0474$\pm$0.0051 & 417.1$\pm$56.2 & $-$0.819$\pm$0.051 & $-$1.714$\pm$0.031 & 1498.7/478 \\
70.591 & 72.794 & 0.0487$\pm$0.0061 & 262.8$\pm$30.7 & $-$0.605$\pm$0.070 & $-$1.650$\pm$0.026 & 3089.7/478 \\
72.794 & 74.372 & 0.0597$\pm$0.0061 & 372.9$\pm$37.6 & $-$0.615$\pm$0.054 & $-$1.772$\pm$0.032 & 1165.1/478 \\
74.372 & 75.362 & 0.0950$\pm$0.0105 & 373.0$\pm$36.6 & $-$0.587$\pm$0.054 & $-$1.874$\pm$0.038 & 1090.3/478 \\
75.362 & 76.443 & 0.1011$\pm$0.0112 & 334.1$\pm$31.7 & $-$0.571$\pm$0.056 & $-$1.879$\pm$0.044 & 725.08/478 \\
76.443 & 78.171 & 0.0513$\pm$0.0051 & 347.2$\pm$37.5 & $-$0.759$\pm$0.053 & $-$1.791$\pm$0.029 & 1247.7/478 \\
78.171 & 86.297 & 0.0109$\pm$0.0011 & 513.8$\pm$78.7 & $-$0.982$\pm$0.043 & $-$1.634$\pm$0.025 & 15102./478 \\
86.297 & 123.477 & 0.0056$\pm$0.0038 & 93.0$\pm$49.1 & $-$0.692$\pm$0.292 & $-$1.312$\pm$0.015 & 15525./478 \\
123.477 & 130.458 & 0.0080$\pm$0.0041 & 143.4$\pm$17.6 & $-$0.864$\pm$0.083 & $-$2.062$\pm$0.134 & 907.00/478 \\
130.458 & 142.336 & $<$0.0018 & 111.8$\pm$53.1 & $-$1.168$\pm$0.212 & $-$1.842$\pm$0.143 & 1020.7/478 \\
\hline
\end{tabular}
\end{table*}

\clearpage

\begin{table*}[!htbp]
\tiny
\caption{BAND Parameters for GRB 100826A. The times $t_\text{start}$ and $t_\text{stop}$ are relative to the GBM trigger time $T_0$.}
\label{tab:taba3}
\begin{tabular}{ccccccc}
\hline\hline
$t_\text{start}$ & $t_\text{stop}$ & $A$ & $E_\text{p}$ & $\alpha$ & $\beta$ & CSTAT/DOF \\
(s) & (s) & (ph~s$^{-1}$~cm$^{-2}$~keV$^{-1}$) & (keV) &  &  & \\
\hline
-2.048 & 9.547 & 0.0551$\pm$0.0145 & 144.3$\pm$14.5 & $-$0.006$\pm$0.160 & $-$1.782$\pm$0.055 & 609.22/356 \\
9.547 & 11.903 & 0.0501$\pm$0.0099 & 296.2$\pm$42.1 & $-$0.545$\pm$0.099 & $-$1.993$\pm$0.100 & 451.52/356 \\
11.903 & 13.740 & 0.0650$\pm$0.0129 & 275.8$\pm$37.3 & $-$0.543$\pm$0.098 & $-$2.032$\pm$0.105 & 412.41/356 \\
13.740 & 14.495 & 0.0781$\pm$0.0159 & 323.8$\pm$51.6 & $-$0.551$\pm$0.100 & $-$1.943$\pm$0.094 & 368.17/356 \\
14.495 & 15.320 & 0.0900$\pm$0.0178 & 312.9$\pm$47.8 & $-$0.551$\pm$0.099 & $-$1.938$\pm$0.082 & 387.25/356 \\
15.320 & 15.979 & 0.0629$\pm$0.0088 & 561.2$\pm$84.0 & $-$0.814$\pm$0.058 & $-$2.813$\pm$0.696 & 349.78/356 \\
15.979 & 16.622 & 0.0550$\pm$0.0075 & 693.2$\pm$127.0 & $-$0.808$\pm$0.061 & $-$2.237$\pm$0.195 & 389.88/356 \\
16.622 & 17.190 & 0.0677$\pm$0.0093 & 551.2$\pm$81.7 & $-$0.747$\pm$0.061 & $-$2.437$\pm$0.275 & 426.53/356 \\
17.190 & 17.712 & 0.1054$\pm$0.0214 & 314.5$\pm$51.2 & $-$0.582$\pm$0.099 & $-$1.891$\pm$0.078 & 359.30/356 \\
17.712 & 18.294 & 0.1062$\pm$0.0207 & 328.0$\pm$47.5 & $-$0.565$\pm$0.092 & $-$2.099$\pm$0.120 & 385.37/356 \\
18.294 & 18.769 & 0.1372$\pm$0.0264 & 328.0$\pm$45.7 & $-$0.485$\pm$0.094 & $-$2.044$\pm$0.099 & 430.26/356 \\
18.769 & 19.175 & 0.0908$\pm$0.0120 & 580.0$\pm$81.2 & $-$0.696$\pm$0.061 & $-$2.502$\pm$0.313 & 397.80/356 \\
19.175 & 19.580 & 0.1196$\pm$0.0191 & 433.9$\pm$65.9 & $-$0.636$\pm$0.077 & $-$2.069$\pm$0.102 & 383.34/356 \\
19.580 & 19.961 & 0.0991$\pm$0.0137 & 581.5$\pm$92.5 & $-$0.681$\pm$0.067 & $-$2.142$\pm$0.130 & 346.77/356 \\
19.961 & 20.408 & 0.1155$\pm$0.0186 & 452.9$\pm$73.0 & $-$0.676$\pm$0.075 & $-$2.070$\pm$0.104 & 359.67/356 \\
20.408 & 20.825 & 0.1587$\pm$0.0276 & 353.2$\pm$45.8 & $-$0.510$\pm$0.085 & $-$2.187$\pm$0.120 & 376.13/356 \\
20.825 & 21.204 & 0.1393$\pm$0.0220 & 424.2$\pm$57.4 & $-$0.560$\pm$0.078 & $-$2.232$\pm$0.141 & 397.45/356 \\
21.204 & 21.603 & 0.1282$\pm$0.0199 & 460.3$\pm$72.3 & $-$0.696$\pm$0.073 & $-$2.130$\pm$0.116 & 375.87/356 \\
21.603 & 22.059 & 0.1219$\pm$0.0193 & 424.2$\pm$63.6 & $-$0.681$\pm$0.074 & $-$2.161$\pm$0.130 & 377.50/356 \\
22.059 & 22.412 & 0.1160$\pm$0.0161 & 539.0$\pm$81.1 & $-$0.657$\pm$0.067 & $-$2.137$\pm$0.119 & 359.72/356 \\
22.412 & 22.834 & 0.1335$\pm$0.0226 & 411.0$\pm$65.6 & $-$0.604$\pm$0.083 & $-$1.952$\pm$0.076 & 391.29/356 \\
22.834 & 23.301 & 0.0928$\pm$0.0140 & 549.7$\pm$104.0 & $-$0.792$\pm$0.067 & $-$1.998$\pm$0.099 & 343.64/356 \\
23.301 & 23.825 & 0.0746$\pm$0.0112 & 554.1$\pm$96.3 & $-$0.756$\pm$0.067 & $-$2.173$\pm$0.162 & 433.43/356 \\
23.825 & 24.471 & 0.0710$\pm$0.0121 & 458.0$\pm$87.6 & $-$0.754$\pm$0.078 & $-$1.938$\pm$0.096 & 402.13/356 \\
24.471 & 25.169 & 0.0824$\pm$0.0159 & 355.3$\pm$61.4 & $-$0.677$\pm$0.088 & $-$1.952$\pm$0.095 & 434.64/356 \\
25.169 & 25.789 & 0.0727$\pm$0.0137 & 399.5$\pm$79.9 & $-$0.716$\pm$0.088 & $-$1.836$\pm$0.077 & 371.05/356 \\
25.789 & 26.336 & 0.0907$\pm$0.0182 & 332.9$\pm$53.9 & $-$0.636$\pm$0.091 & $-$2.004$\pm$0.119 & 424.82/356 \\
26.336 & 27.189 & 0.0856$\pm$0.0168 & 309.3$\pm$48.8 & $-$0.710$\pm$0.088 & $-$2.060$\pm$0.120 & 433.55/356 \\
27.189 & 28.154 & 0.0928$\pm$0.0210 & 220.0$\pm$33.9 & $-$0.564$\pm$0.115 & $-$1.895$\pm$0.093 & 364.10/356 \\
28.154 & 28.989 & 0.1055$\pm$0.0222 & 266.2$\pm$37.2 & $-$0.628$\pm$0.094 & $-$2.128$\pm$0.124 & 438.88/356 \\
28.989 & 29.888 & 0.0709$\pm$0.0148 & 316.4$\pm$57.6 & $-$0.830$\pm$0.086 & $-$2.067$\pm$0.133 & 427.39/356 \\
29.888 & 31.139 & 0.0452$\pm$0.0098 & 336.6$\pm$75.2 & $-$0.937$\pm$0.086 & $-$1.991$\pm$0.144 & 394.69/356 \\
31.139 & 32.985 & 0.0408$\pm$0.0092 & 262.2$\pm$53.6 & $-$0.812$\pm$0.102 & $-$1.837$\pm$0.092 & 399.37/356 \\
32.985 & 35.735 & 0.0365$\pm$0.0085 & 228.4$\pm$37.4 & $-$0.723$\pm$0.103 & $-$1.956$\pm$0.126 & 387.28/356 \\
35.735 & 37.794 & 0.0868$\pm$0.0321 & 126.6$\pm$21.1 & $-$0.321$\pm$0.199 & $-$1.765$\pm$0.071 & 393.57/356 \\
37.794 & 40.792 & 0.0496$\pm$0.0189 & 121.6$\pm$25.3 & $-$0.542$\pm$0.195 & $-$1.722$\pm$0.068 & 407.95/356 \\
40.792 & 46.181 & 0.0710$\pm$0.0216 & 117.2$\pm$15.9 & $-$0.526$\pm$0.154 & $-$1.963$\pm$0.094 & 424.55/356 \\
46.181 & 60.975 & 0.0493$\pm$0.0150 & 108.4$\pm$13.3 & $-$0.328$\pm$0.159 & $-$1.777$\pm$0.051 & 520.17/356 \\
60.975 & 62.775 & 0.0614$\pm$0.0135 & 232.6$\pm$35.3 & $-$0.768$\pm$0.094 & $-$2.094$\pm$0.147 & 442.50/356 \\
62.775 & 64.610 & 0.0456$\pm$0.0105 & 258.6$\pm$45.7 & $-$0.951$\pm$0.082 & $-$2.221$\pm$0.236 & 402.38/356 \\
64.610 & 70.330 & 0.0254$\pm$0.0057 & 211.1$\pm$35.6 & $-$0.917$\pm$0.088 & $-$2.016$\pm$0.150 & 426.15/356 \\
70.330 & 72.515 & 0.0279$\pm$0.0041 & 549.0$\pm$85.8 & $-$0.968$\pm$0.051 & $<-$12.27 & 410.69/356 \\
72.515 & 73.736 & 0.0520$\pm$0.0106 & 346.4$\pm$73.5 & $-$0.941$\pm$0.080 & $-$2.020$\pm$0.136 & 395.10/356 \\
73.736 & 76.263 & 0.0455$\pm$0.0106 & 209.4$\pm$31.9 & $-$0.813$\pm$0.094 & $-$2.101$\pm$0.170 & 375.69/356 \\
76.263 & 78.063 & 0.0561$\pm$0.0134 & 231.3$\pm$34.0 & $-$0.889$\pm$0.083 & $-$2.472$\pm$0.365 & 377.34/356 \\
78.063 & 80.235 & 0.0354$\pm$0.0087 & 263.8$\pm$56.3 & $-$1.199$\pm$0.070 & $-$2.476$\pm$0.547 & 392.60/356 \\
80.235 & 83.975 & 0.1548$\pm$0.2100 & 45.8$\pm$12.1 & $-$0.143$\pm$0.553 & $-$1.672$\pm$0.042 & 412.44/356 \\
83.975 & 89.057 & 0.1167$\pm$0.1011 & 51.3$\pm$10.5 & $-$0.384$\pm$0.364 & $-$1.805$\pm$0.049 & 386.98/356 \\
89.057 & 90.011 & 0.0838$\pm$0.0258 & 169.4$\pm$21.9 & $-$1.043$\pm$0.082 & $-$3.072$\pm$1.090 & 380.52/356 \\
90.011 & 91.665 & 0.1179$\pm$0.0450 & 82.9$\pm$11.8 & $-$0.950$\pm$0.151 & $-$2.301$\pm$0.192 & 400.87/356 \\
91.665 & 98.549 & 0.0659$\pm$0.0566 & 46.9$\pm$9.6 & $-$0.552$\pm$0.343 & $-$1.857$\pm$0.050 & 440.67/356 \\
98.549 & 121.856 & $<$1.5170 & 29.0$\pm$6.8 & $<-$1.102 & $-$1.791$\pm$0.053 & 581.38/356 \\
\hline
\end{tabular}
\end{table*}

\clearpage

\begin{table*}[!htbp]
\tiny
\caption{BAND Parameters for GRB 101123A. The times $t_\text{start}$ and $t_\text{stop}$ are relative to the GBM trigger time $T_0$.}
\label{tab:taba4}
\begin{tabular}{ccccccc}
\hline\hline
$t_\text{start}$ & $t_\text{stop}$ & $A$ & $E_\text{p}$ & $\alpha$ & $\beta$ & CSTAT/DOF \\
(s) & (s) & (ph~s$^{-1}$~cm$^{-2}$~keV$^{-1}$) & (keV) &  &  & \\
\hline
38.912 & 43.589 & 0.0179$\pm$0.0033 & 497.5$\pm$98.7 & $-$0.549$\pm$0.101 & $-$1.926$\pm$0.100 & 447.95/356 \\
43.589 & 44.391 & 0.0380$\pm$0.0033 & 1635.0$\pm$295.0 & $-$0.717$\pm$0.048 & $-$2.026$\pm$0.114 & 370.62/356 \\
44.391 & 44.844 & 0.0770$\pm$0.0104 & 683.7$\pm$129.0 & $-$0.583$\pm$0.073 & $-$1.753$\pm$0.057 & 380.88/356 \\
44.844 & 45.250 & 0.1016$\pm$0.0166 & 506.6$\pm$90.5 & $-$0.606$\pm$0.077 & $-$1.884$\pm$0.075 & 372.85/356 \\
45.250 & 45.628 & 0.1530$\pm$0.0276 & 374.4$\pm$54.5 & $-$0.486$\pm$0.088 & $-$1.989$\pm$0.086 & 345.89/356 \\
45.628 & 45.949 & 0.1407$\pm$0.0267 & 380.9$\pm$67.5 & $-$0.541$\pm$0.093 & $-$1.777$\pm$0.060 & 372.09/356 \\
45.949 & 46.255 & 0.1001$\pm$0.0148 & 577.7$\pm$98.7 & $-$0.728$\pm$0.064 & $-$2.132$\pm$0.150 & 399.98/356 \\
46.255 & 46.552 & 0.1173$\pm$0.0179 & 544.2$\pm$94.3 & $-$0.726$\pm$0.067 & $-$2.088$\pm$0.125 & 366.58/356 \\
46.552 & 46.836 & 0.1243$\pm$0.0206 & 467.8$\pm$75.3 & $-$0.605$\pm$0.076 & $-$2.044$\pm$0.118 & 414.08/356 \\
46.836 & 47.262 & 0.1339$\pm$0.0303 & 278.5$\pm$47.1 & $-$0.702$\pm$0.095 & $-$2.040$\pm$0.128 & 404.28/356 \\
47.262 & 47.831 & 0.0808$\pm$0.0155 & 416.5$\pm$84.7 & $-$0.825$\pm$0.077 & $-$2.061$\pm$0.144 & 337.31/356 \\
47.831 & 48.538 & 0.0739$\pm$0.0137 & 423.0$\pm$84.1 & $-$0.826$\pm$0.076 & $-$2.059$\pm$0.134 & 374.63/356 \\
48.538 & 49.070 & 0.0923$\pm$0.0173 & 388.3$\pm$61.6 & $-$0.679$\pm$0.078 & $-$2.255$\pm$0.207 & 382.10/356 \\
49.070 & 49.701 & 0.1666$\pm$0.0424 & 198.2$\pm$25.3 & $-$0.482$\pm$0.114 & $-$2.196$\pm$0.167 & 396.86/356 \\
49.701 & 50.067 & 0.1229$\pm$0.0174 & 453.2$\pm$50.6 & $-$0.515$\pm$0.068 & $-$2.815$\pm$0.522 & 368.62/356 \\
50.067 & 50.332 & 0.1673$\pm$0.0188 & 533.5$\pm$49.6 & $-$0.582$\pm$0.055 & $-$3.884$\pm$2.3 & 364.11/356 \\
50.332 & 50.600 & 0.1184$\pm$0.0151 & 616.6$\pm$83.3 & $-$0.711$\pm$0.056 & $-$2.630$\pm$0.367 & 394.00/356 \\
50.600 & 51.027 & 0.1445$\pm$0.0320 & 267.1$\pm$41.8 & $-$0.662$\pm$0.095 & $-$2.042$\pm$0.117 & 383.01/356 \\
51.027 & 51.565 & 0.0824$\pm$0.0136 & 443.1$\pm$68.5 & $-$0.915$\pm$0.059 & $-$3.471$\pm$2.750 & 368.09/356 \\
51.565 & 51.904 & 0.1089$\pm$0.0123 & 648.0$\pm$74.4 & $-$0.770$\pm$0.049 & $<-$9.740 & 348.95/356 \\
51.904 & 52.133 & 0.1842$\pm$0.0233 & 457.0$\pm$42.2 & $-$0.659$\pm$0.055 & $<-$14.27 & 331.89/356 \\
52.133 & 52.444 & 0.1727$\pm$0.0318 & 374.4$\pm$56.8 & $-$0.641$\pm$0.079 & $-$2.204$\pm$0.145 & 412.50/356 \\
52.444 & 52.701 & 0.1704$\pm$0.0201 & 518.0$\pm$51.0 & $-$0.705$\pm$0.052 & $<-$14.90 & 394.79/356 \\
52.701 & 53.008 & 0.1605$\pm$0.0253 & 420.4$\pm$56.2 & $-$0.672$\pm$0.069 & $-$2.523$\pm$0.283 & 437.42/356 \\
53.008 & 53.355 & 0.0999$\pm$0.0127 & 565.9$\pm$68.1 & $-$0.737$\pm$0.054 & $-$4.167$\pm$5.450 & 401.64/356 \\
53.355 & 53.681 & 0.1462$\pm$0.0230 & 425.7$\pm$54.1 & $-$0.617$\pm$0.068 & $-$2.588$\pm$0.329 & 333.49/356 \\
53.681 & 54.170 & 0.0849$\pm$0.0157 & 410.1$\pm$71.3 & $-$0.823$\pm$0.071 & $-$2.294$\pm$0.240 & 377.17/356 \\
54.170 & 55.513 & 0.0225$\pm$0.0047 & 688.9$\pm$240.0 & $-$1.053$\pm$0.068 & $-$2.049$\pm$0.248 & 394.51/356 \\
55.513 & 55.962 & 0.1402$\pm$0.0264 & 346.1$\pm$51.1 & $-$0.663$\pm$0.081 & $-$2.230$\pm$0.161 & 419.37/356 \\
55.962 & 56.334 & 0.0982$\pm$0.0124 & 559.1$\pm$60.4 & $-$0.630$\pm$0.057 & $-$7.559$\pm$264.0 & 357.26/356 \\
56.334 & 56.672 & 0.1198$\pm$0.0194 & 431.7$\pm$60.8 & $-$0.708$\pm$0.067 & $-$2.483$\pm$0.324 & 346.27/356 \\
56.672 & 58.101 & 0.0776$\pm$0.0235 & 199.9$\pm$32.8 & $-$0.972$\pm$0.093 & $-$2.768$\pm$0.627 & 430.97/356 \\
58.101 & 62.873 & 0.0291$\pm$0.0105 & 171.3$\pm$42.3 & $-$0.865$\pm$0.149 & $-$1.996$\pm$0.187 & 449.63/356 \\
62.873 & 67.584 & 0.2631$\pm$1.4600 & 68.4$\pm$43.5 & $<-$1.497 & $-$1.418$\pm$0.111 & 489.31/356 \\
81.920 & 89.903 & 0.0113$\pm$0.0025 & 568.6$\pm$175.0 & $-$1.146$\pm$0.069 & $<-$17.00 & 518.16/356 \\
89.903 & 92.126 & 0.0323$\pm$0.0090 & 288.3$\pm$53.7 & $-$0.946$\pm$0.084 & $-$3.685$\pm$5.650 & 419.32/356 \\
92.126 & 93.570 & 0.0327$\pm$0.0095 & 283.1$\pm$62.0 & $-$0.968$\pm$0.090 & $-$2.350$\pm$0.504 & 466.08/356 \\
93.570 & 94.965 & 0.0610$\pm$0.0217 & 140.1$\pm$31.2 & $-$0.902$\pm$0.146 & $-$2.005$\pm$0.173 & 360.96/356 \\
94.965 & 100.352 & 0.0131$\pm$0.0087 & 109.2$\pm$50.4 & $-$0.925$\pm$0.281 & $-$1.778$\pm$0.142 & 441.31/356 \\
140.288 & 143.695 & 0.0172$\pm$0.0051 & 407.7$\pm$149.0 & $-$1.208$\pm$0.080 & $-$2.602$\pm$1.250 & 373.85/356 \\
143.695 & 145.029 & 0.0200$\pm$0.0050 & 537.8$\pm$161.0 & $-$1.095$\pm$0.069 & $-$2.940$\pm$2.800 & 388.75/356 \\
145.029 & 146.541 & 0.0360$\pm$0.0137 & 158.5$\pm$42.3 & $-$0.944$\pm$0.142 & $-$1.936$\pm$0.187 & 375.53/356 \\
146.541 & 149.288 & 0.0444$\pm$0.0203 & 109.1$\pm$26.4 & $-$0.847$\pm$0.192 & $-$1.997$\pm$0.176 & 393.57/356 \\
149.288 & 155.648 & $<$0.0793 & 55.3$\pm$19.2 & $-$0.435$\pm$0.982 & $-$1.746$\pm$0.132 & 365.35/356 \\
\hline
\end{tabular}
\end{table*}

\clearpage

\begin{table*}[!htbp]
\tiny
\caption{BAND Parameters for GRB 120526A. The times $t_\text{start}$ and $t_\text{stop}$ are relative to the GBM trigger time $T_0$.}
\label{tab:taba5}
\begin{tabular}{ccccccc}
\hline\hline
$t_\text{start}$ & $t_\text{stop}$ & $A$ & $E_\text{p}$ & $\alpha$ & $\beta$ & CSTAT/DOF \\
(s) & (s) & (ph~s$^{-1}$~cm$^{-2}$~keV$^{-1}$) & (keV) &  &  & \\
\hline
-2.048 & 3.367 & 0.0144$\pm$0.0023 & 625.6$\pm$100.0 & $-$0.610$\pm$0.145 & $<-$8.973 & 277.98/239 \\
3.367 & 5.382 & 0.0430$\pm$0.0069 & 457.2$\pm$57.8 & $-$0.647$\pm$0.128 & $<-$12.32 & 248.49/239 \\
5.382 & 7.254 & 0.0339$\pm$0.0041 & 720.7$\pm$79.8 & $-$0.533$\pm$0.122 & $<-$10.83 & 256.79/239 \\
7.254 & 8.973 & 0.0262$\pm$0.0030 & 1389.0$\pm$221.0 & $-$0.893$\pm$0.083 & $<-$6.327 & 302.93/239 \\
8.973 & 11.365 & 0.0214$\pm$0.0032 & 865.8$\pm$203.0 & $-$0.954$\pm$0.106 & $<-$5.560 & 288.98/239 \\
11.365 & 13.269 & 0.0340$\pm$0.0042 & 711.4$\pm$88.2 & $-$0.630$\pm$0.118 & $<-$11.92 & 236.42/239 \\
13.269 & 15.565 & 0.0224$\pm$0.0027 & 954.3$\pm$144.0 & $-$0.521$\pm$0.129 & $-$2.691$\pm$0.380 & 271.67/239 \\
15.565 & 16.889 & 0.0220$\pm$0.0029 & 1463.0$\pm$369.0 & $-$0.724$\pm$0.113 & $-$2.174$\pm$0.194 & 263.50/239 \\
16.889 & 18.729 & 0.0288$\pm$0.0034 & 1224.0$\pm$198.0 & $-$0.894$\pm$0.090 & $<-$6.337 & 264.96/239 \\
18.729 & 20.182 & 0.0334$\pm$0.0042 & 1100.0$\pm$236.0 & $-$0.911$\pm$0.098 & $-$2.885$\pm$0.660 & 248.03/239 \\
20.182 & 22.412 & 0.0302$\pm$0.0044 & 683.2$\pm$132.0 & $-$0.693$\pm$0.134 & $-$2.410$\pm$0.271 & 214.16/239 \\
22.412 & 24.484 & 0.0214$\pm$0.0028 & 984.4$\pm$163.0 & $-$0.698$\pm$0.117 & $<-$6.364 & 241.56/239 \\
24.484 & 26.325 & 0.0300$\pm$0.0037 & 936.3$\pm$158.0 & $-$0.768$\pm$0.109 & $-$3.186$\pm$0.976 & 234.93/239 \\
26.325 & 28.016 & 0.0331$\pm$0.0042 & 781.7$\pm$114.0 & $-$0.743$\pm$0.108 & $<-$11.92 & 247.35/239 \\
28.016 & 29.555 & 0.0278$\pm$0.0035 & 1119.0$\pm$238.0 & $-$0.791$\pm$0.108 & $-$2.533$\pm$0.347 & 285.22/239 \\
29.555 & 30.980 & 0.0337$\pm$0.0052 & 640.2$\pm$109.0 & $-$0.653$\pm$0.129 & $-$3.159$\pm$1.240 & 257.64/239 \\
30.980 & 33.168 & 0.0298$\pm$0.0039 & 746.7$\pm$117.0 & $-$0.695$\pm$0.125 & $-$3.105$\pm$0.903 & 285.57/239 \\
33.168 & 34.993 & 0.0280$\pm$0.0042 & 853.7$\pm$219.0 & $-$0.918$\pm$0.112 & $-$2.499$\pm$0.407 & 246.47/239 \\
34.993 & 36.553 & 0.0366$\pm$0.0062 & 697.4$\pm$204.0 & $-$0.994$\pm$0.115 & $-$2.269$\pm$0.253 & 255.41/239 \\
36.553 & 37.691 & 0.0362$\pm$0.0079 & 475.6$\pm$109.0 & $-$0.991$\pm$0.120 & $<-$9.537 & 246.99/239 \\
37.691 & 39.667 & 0.0253$\pm$0.0037 & 821.6$\pm$166.0 & $-$0.825$\pm$0.115 & $-$3.497$\pm$2.140 & 215.69/239 \\
39.667 & 41.306 & 0.0374$\pm$0.0061 & 532.0$\pm$86.7 & $-$0.813$\pm$0.119 & $<-$14.64 & 242.91/239 \\
41.306 & 42.977 & 0.0404$\pm$0.0074 & 466.1$\pm$77.8 & $-$0.757$\pm$0.131 & $-$3.940$\pm$4.880 & 204.96/239 \\
42.977 & 44.596 & 0.0452$\pm$0.0110 & 342.3$\pm$56.5 & $-$0.771$\pm$0.141 & $<-$5.606 & 240.63/239 \\
44.596 & 46.530 & 0.0407$\pm$0.0083 & 395.0$\pm$62.2 & $-$0.781$\pm$0.132 & $<-$14.90 & 253.36/239 \\
46.530 & 56.166 & 0.0064$\pm$0.0047 & $<$227.1 & $-$1.568$\pm$0.307 & $-$1.966$\pm$0.374 & 298.50/239 \\
56.166 & 67.584 & $<$0.0052 & $<$21.2 & $<-$1.847 & $-$1.963$\pm$0.212 & 301.17/239 \\
\hline
\end{tabular}
\end{table*}

\clearpage

\begin{table*}[!htbp]
\tiny
\caption{BAND Parameters for GRB 130427A. The times $t_\text{start}$ and $t_\text{stop}$ are relative to the GBM trigger time $T_0$.}
\label{tab:taba6}
\begin{tabular}{ccccccc}
\hline\hline
$t_\text{start}$ & $t_\text{stop}$ & $A$ & $E_\text{p}$ & $\alpha$ & $\beta$ & CSTAT/DOF \\
(s) & (s) & (ph~s$^{-1}$~cm$^{-2}$~keV$^{-1}$) & (keV) &  &  & \\
\hline
-0.064 & 0.145 & 0.0821$\pm$0.0080 & 1246.0$\pm$126.0 & $-$0.386$\pm$0.081 & $-$2.916$\pm$0.359 & 1489.3/357 \\
0.145 & 0.214 & 0.1936$\pm$0.0197 & 918.8$\pm$73.1 & $-$0.116$\pm$0.089 & $-$3.719$\pm$0.799 & 1275.0/357 \\
0.214 & 0.268 & 0.3744$\pm$0.0374 & 764.6$\pm$62.0 & $-$0.363$\pm$0.074 & $-$4.031$\pm$1.280 & 1246.8/357 \\
0.268 & 0.313 & 0.4275$\pm$0.0506 & 657.9$\pm$63.1 & $-$0.319$\pm$0.085 & $-$3.400$\pm$0.702 & 1168.7/357 \\
0.313 & 0.358 & 0.5104$\pm$0.0558 & 569.5$\pm$42.7 & $-$0.332$\pm$0.078 & $-$4.889$\pm$3.570 & 1163.6/357 \\
0.358 & 0.398 & 0.4528$\pm$0.0510 & 734.1$\pm$75.3 & $-$0.560$\pm$0.070 & $-$3.750$\pm$1.090 & 1169.3/357 \\
0.398 & 0.434 & 0.7357$\pm$0.1020 & 486.6$\pm$47.5 & $-$0.593$\pm$0.075 & $-$3.979$\pm$1.580 & 1115.3/357 \\
0.434 & 0.472 & 0.5783$\pm$0.0660 & 617.0$\pm$60.7 & $-$0.610$\pm$0.066 & $<-$11.92 & 1119.4/357 \\
0.472 & 0.505 & 0.5876$\pm$0.0804 & 572.7$\pm$69.1 & $-$0.676$\pm$0.072 & $-$3.286$\pm$0.834 & 1090.3/357 \\
0.505 & 0.541 & 0.9841$\pm$0.2210 & 319.3$\pm$42.3 & $-$0.466$\pm$0.111 & $-$2.721$\pm$0.281 & 1098.2/357 \\
0.541 & 0.575 & 0.6740$\pm$0.1011 & 465.0$\pm$53.0 & $-$0.647$\pm$0.077 & $-$3.512$\pm$1.050 & 1100.0/357 \\
0.575 & 0.614 & 0.8132$\pm$0.1500 & 379.0$\pm$47.8 & $-$0.593$\pm$0.089 & $-$2.949$\pm$0.408 & 1113.6/357 \\
0.614 & 0.656 & 0.4534$\pm$0.0718 & 505.2$\pm$65.4 & $-$0.800$\pm$0.070 & $-$3.815$\pm$2.120 & 1107.9/357 \\
0.656 & 0.696 & 0.5599$\pm$0.0885 & 423.0$\pm$44.6 & $-$0.726$\pm$0.072 & $<-$6.620 & 1089.5/357 \\
0.696 & 0.736 & 0.5940$\pm$0.1160 & 366.3$\pm$50.1 & $-$0.762$\pm$0.087 & $-$3.075$\pm$0.656 & 1123.8/357 \\
0.736 & 0.780 & 0.4775$\pm$0.1011 & 383.8$\pm$65.1 & $-$0.746$\pm$0.095 & $-$2.593$\pm$0.350 & 1170.2/357 \\
0.780 & 0.824 & 0.6696$\pm$0.1150 & 339.6$\pm$32.5 & $-$0.732$\pm$0.077 & $<-$9.537 & 1075.6/357 \\
0.824 & 0.870 & 0.4616$\pm$0.0926 & 348.3$\pm$43.1 & $-$0.813$\pm$0.078 & $<-$9.537 & 1040.5/357 \\
0.870 & 0.927 & 0.8204$\pm$0.2170 & 244.9$\pm$32.2 & $-$0.686$\pm$0.108 & $-$3.027$\pm$0.448 & 1008.3/357 \\
0.927 & 0.994 & 0.6554$\pm$0.1690 & 234.1$\pm$28.2 & $-$0.696$\pm$0.104 & $-$3.118$\pm$0.554 & 1156.6/357 \\
0.994 & 1.064 & 0.5467$\pm$0.1660 & 210.9$\pm$33.9 & $-$0.732$\pm$0.132 & $-$2.504$\pm$0.252 & 1107.2/357 \\
1.064 & 1.147 & 0.4542$\pm$0.1370 & 204.3$\pm$31.1 & $-$0.766$\pm$0.126 & $-$2.591$\pm$0.311 & 1168.0/357 \\
1.147 & 1.226 & 0.9292$\pm$0.3080 & 162.6$\pm$19.5 & $-$0.489$\pm$0.154 & $-$2.652$\pm$0.261 & 1155.7/357 \\
1.226 & 1.300 & 0.9699$\pm$0.2980 & 169.2$\pm$13.3 & $-$0.577$\pm$0.109 & $<-$7.937 & 1052.7/357 \\
1.300 & 1.378 & 0.8825$\pm$0.2450 & 182.0$\pm$14.9 & $-$0.660$\pm$0.100 & $<-$5.563 & 1102.9/357 \\
1.378 & 1.447 & 1.3581$\pm$0.4850 & 144.9$\pm$13.9 & $-$0.400$\pm$0.150 & $-$3.036$\pm$0.384 & 1034.1/357 \\
1.447 & 1.529 & 0.5947$\pm$0.1870 & 176.2$\pm$18.9 & $-$0.663$\pm$0.123 & $-$3.338$\pm$0.738 & 1125.5/357 \\
1.529 & 1.619 & 0.5603$\pm$0.1870 & 170.0$\pm$15.6 & $-$0.670$\pm$0.112 & $<-$4.840 & 1127.0/357 \\
1.619 & 1.708 & 0.8494$\pm$0.3490 & 131.1$\pm$10.0 & $-$0.591$\pm$0.126 & $-$5.137$\pm$5.110 & 1071.8/357 \\
1.708 & 1.795 & 0.5043$\pm$0.1810 & 148.3$\pm$16.4 & $-$0.632$\pm$0.140 & $-$2.950$\pm$0.470 & 1144.4/357 \\
1.795 & 1.903 & 1.0900$\pm$0.4041 & 129.2$\pm$10.3 & $-$0.493$\pm$0.137 & $-$3.475$\pm$0.607 & 1103.9/357 \\
1.903 & 2.002 & 0.6567$\pm$0.2420 & 133.3$\pm$13.8 & $-$0.613$\pm$0.143 & $-$2.994$\pm$0.482 & 1120.9/357 \\
2.002 & 2.123 & 0.8216$\pm$0.3150 & 123.5$\pm$11.7 & $-$0.600$\pm$0.143 & $-$3.164$\pm$0.519 & 1116.5/357 \\
2.123 & 2.249 & 0.7220$\pm$0.3370 & 102.8$\pm$12.0 & $-$0.344$\pm$0.218 & $-$2.675$\pm$0.320 & 1145.5/357 \\
2.249 & 2.405 & 0.5713$\pm$0.2600 & 119.9$\pm$9.0 & $-$0.660$\pm$0.128 & $<-$6.759 & 1165.1/357 \\
2.405 & 2.613 & 0.7771$\pm$0.3810 & 99.5$\pm$7.5 & $-$0.748$\pm$0.143 & $-$4.253$\pm$1.560 & 1255.3/357 \\
2.613 & 2.752 & $<$1.4110 & 87.3$\pm$6.5 & $-$0.302$\pm$0.222 & $-$4.909$\pm$3.920 & 1049.3/357 \\
\hline
\end{tabular}
\end{table*}

\clearpage

\begin{table*}[!htbp]
\tiny
\caption{BAND Parameters for GRB 130504C. The times $t_\text{start}$ and $t_\text{stop}$ are relative to the GBM trigger time $T_0$.}
\label{tab:taba7}
\begin{tabular}{ccccccc}
\hline\hline
$t_\text{start}$ & $t_\text{stop}$ & $A$ & $E_\text{p}$ & $\alpha$ & $\beta$ & CSTAT/DOF \\
(s) & (s) & (ph~s$^{-1}$~cm$^{-2}$~keV$^{-1}$) & (keV) &  &  & \\
\hline
-6.144 & 14.601 & 0.0074$\pm$0.0008 & 969.0$\pm$180.0 & $-$0.892$\pm$0.049 & $-$2.236$\pm$0.208 & 593.56/359 \\
14.601 & 15.059 & 0.0640$\pm$0.0071 & 952.5$\pm$164.0 & $-$0.784$\pm$0.057 & $-$2.420$\pm$0.259 & 388.66/359 \\
15.059 & 15.467 & 0.0539$\pm$0.0049 & 2096.0$\pm$454.0 & $-$0.945$\pm$0.041 & $-$2.312$\pm$0.261 & 383.17/359 \\
15.467 & 15.851 & 0.1054$\pm$0.0126 & 681.7$\pm$103.0 & $-$0.732$\pm$0.064 & $-$2.311$\pm$0.169 & 368.60/359 \\
15.851 & 16.140 & 0.1040$\pm$0.0108 & 959.4$\pm$158.0 & $-$0.824$\pm$0.052 & $-$2.388$\pm$0.220 & 372.68/359 \\
16.140 & 16.485 & 0.1134$\pm$0.0170 & 565.5$\pm$101.0 & $-$0.757$\pm$0.073 & $-$2.122$\pm$0.131 & 358.14/359 \\
16.485 & 16.926 & 0.0876$\pm$0.0120 & 596.3$\pm$104.0 & $-$0.813$\pm$0.067 & $-$2.210$\pm$0.177 & 414.67/359 \\
16.926 & 17.368 & 0.0700$\pm$0.0087 & 727.6$\pm$114.0 & $-$0.911$\pm$0.053 & $-$3.828$\pm$3.170 & 418.61/359 \\
17.368 & 18.042 & 0.0528$\pm$0.0080 & 586.7$\pm$112.0 & $-$0.853$\pm$0.068 & $-$2.246$\pm$0.232 & 459.53/359 \\
18.042 & 18.791 & 0.0594$\pm$0.0103 & 522.2$\pm$120.0 & $-$1.011$\pm$0.068 & $-$2.189$\pm$0.217 & 380.92/359 \\
18.791 & 20.215 & 0.0354$\pm$0.0067 & 487.2$\pm$121.0 & $-$1.025$\pm$0.073 & $-$2.169$\pm$0.238 & 340.56/359 \\
20.215 & 24.224 & 0.0283$\pm$0.0067 & 245.2$\pm$49.0 & $-$0.834$\pm$0.109 & $-$1.954$\pm$0.130 & 437.97/359 \\
24.224 & 25.900 & 0.0468$\pm$0.0112 & 266.8$\pm$55.3 & $-$0.933$\pm$0.100 & $-$2.079$\pm$0.165 & 422.51/359 \\
25.900 & 27.775 & 0.0189$\pm$0.0056 & 300.3$\pm$105.0 & $-$1.204$\pm$0.096 & $-$2.016$\pm$0.276 & 399.99/359 \\
27.775 & 28.549 & 0.0482$\pm$0.0089 & 517.4$\pm$124.0 & $-$1.142$\pm$0.063 & $-$2.707$\pm$0.891 & 449.85/359 \\
28.549 & 28.831 & 0.1072$\pm$0.0143 & 793.9$\pm$190.0 & $-$1.011$\pm$0.058 & $-$2.103$\pm$0.151 & 420.52/359 \\
28.831 & 29.323 & 0.0781$\pm$0.0139 & 538.3$\pm$141.0 & $-$0.999$\pm$0.074 & $-$2.022$\pm$0.130 & 428.27/359 \\
29.323 & 30.661 & 0.0274$\pm$0.0046 & 758.2$\pm$222.0 & $-$1.133$\pm$0.059 & $-$2.451$\pm$0.566 & 433.24/359 \\
30.661 & 30.978 & 0.1087$\pm$0.0164 & 552.3$\pm$93.0 & $-$0.686$\pm$0.076 & $-$2.160$\pm$0.147 & 399.00/359 \\
30.978 & 31.173 & 0.1219$\pm$0.0150 & 951.2$\pm$196.0 & $-$0.818$\pm$0.059 & $-$2.054$\pm$0.124 & 354.79/359 \\
31.173 & 31.397 & 0.1399$\pm$0.0207 & 640.3$\pm$131.0 & $-$0.818$\pm$0.070 & $-$2.024$\pm$0.106 & 404.81/359 \\
31.397 & 31.654 & 0.1717$\pm$0.0338 & 376.4$\pm$72.8 & $-$0.852$\pm$0.085 & $-$2.142$\pm$0.148 & 356.74/359 \\
31.654 & 32.002 & 0.1825$\pm$0.0460 & 237.5$\pm$44.7 & $-$0.766$\pm$0.117 & $-$2.022$\pm$0.117 & 389.46/359 \\
32.002 & 32.962 & 0.0626$\pm$0.0172 & 217.8$\pm$66.2 & $-$0.990$\pm$0.132 & $-$1.812$\pm$0.095 & 440.69/359 \\
32.962 & 49.001 & 0.0522$\pm$0.0785 & 44.9$\pm$17.8 & $-$0.553$\pm$0.602 & $-$1.718$\pm$0.044 & 784.55/359 \\
49.001 & 51.264 & 0.1290$\pm$0.1560 & 66.4$\pm$23.7 & $-$0.234$\pm$0.542 & $-$1.584$\pm$0.046 & 382.04/359 \\
51.264 & 52.063 & 0.6384$\pm$1.0200 & 53.7$\pm$15.7 & $-$0.189$\pm$0.688 & $-$1.603$\pm$0.046 & 388.96/359 \\
52.063 & 63.008 & $<$1.0260 & 29.5$\pm$8.3 & $-$0.475$\pm$1.250 & $-$1.771$\pm$0.042 & 670.35/359 \\
63.008 & 64.051 & 0.0517$\pm$0.0100 & 429.3$\pm$88.0 & $-$1.060$\pm$0.069 & $-$2.845$\pm$1.050 & 407.86/359 \\
64.051 & 64.571 & 0.0582$\pm$0.0099 & 693.2$\pm$196.0 & $-$1.172$\pm$0.058 & $-$2.556$\pm$0.564 & 393.54/359 \\
64.571 & 66.126 & 0.0371$\pm$0.0088 & 349.0$\pm$75.8 & $-$1.232$\pm$0.067 & $<-$5.742 & 423.05/359 \\
66.126 & 67.813 & 0.0287$\pm$0.0073 & 344.8$\pm$84.0 & $-$1.137$\pm$0.079 & $-$2.623$\pm$0.944 & 399.31/359 \\
67.813 & 69.496 & 0.0292$\pm$0.0063 & 463.6$\pm$138.0 & $-$1.182$\pm$0.073 & $-$2.324$\pm$0.441 & 467.87/359 \\
69.496 & 69.836 & 0.0896$\pm$0.0105 & 725.0$\pm$88.0 & $-$0.743$\pm$0.056 & $<-$8.680 & 372.08/359 \\
69.836 & 70.212 & 0.1048$\pm$0.0141 & 553.5$\pm$73.1 & $-$0.834$\pm$0.060 & $-$3.928$\pm$3.460 & 369.74/359 \\
70.212 & 70.748 & 0.1316$\pm$0.0328 & 243.4$\pm$40.1 & $-$0.724$\pm$0.112 & $-$2.167$\pm$0.156 & 363.61/359 \\
70.748 & 76.011 & 0.0163$\pm$0.0054 & 235.3$\pm$43.6 & $-$1.217$\pm$0.078 & $<-$7.767 & 557.39/359 \\
76.011 & 78.210 & 0.0354$\pm$0.0115 & 160.4$\pm$51.4 & $-$1.229$\pm$0.130 & $-$2.021$\pm$0.185 & 373.22/359 \\
78.210 & 121.856 &$<$ 0.0177 & $<$35.0 & $-$0.898$\pm$1.300 & $-$1.684$\pm$0.040 & 1483.3/359 \\
\hline
\end{tabular}
\end{table*}

\clearpage

\begin{table*}[!htbp]
\tiny
\caption{BAND Parameters for GRB 130606B. The times $t_\text{start}$ and $t_\text{stop}$ are relative to the GBM trigger time $T_0$.}
\label{tab:taba8}
\begin{tabular}{ccccccc}
\hline\hline
$t_\text{start}$ & $t_\text{stop}$ & $A$ & $E_\text{p}$ & $\alpha$ & $\beta$ & CSTAT/DOF \\
(s) & (s) & (ph~s$^{-1}$~cm$^{-2}$~keV$^{-1}$) & (keV) &  &  & \\
\hline
-3.072 & 8.039 & 0.0406$\pm$0.0034 & 314.8$\pm$19.8 & $-$0.561$\pm$0.046 & $-$2.012$\pm$0.048 & 592.26/476 \\
8.039 & 9.019 & 0.1460$\pm$0.0139 & 370.0$\pm$31.6 & $-$0.730$\pm$0.044 & $-$2.092$\pm$0.064 & 515.85/476 \\
9.019 & 9.424 & 0.1393$\pm$0.0070 & 1951.0$\pm$234.0 & $-$1.011$\pm$0.021 & $-$2.173$\pm$0.107 & 569.35/476 \\
9.424 & 9.815 & 0.1949$\pm$0.0139 & 699.7$\pm$72.4 & $-$0.846$\pm$0.033 & $-$2.018$\pm$0.062 & 569.74/476 \\
9.815 & 10.220 & 0.2004$\pm$0.0145 & 672.6$\pm$69.5 & $-$0.846$\pm$0.033 & $-$2.005$\pm$0.059 & 539.28/476 \\
10.220 & 10.764 & 0.1765$\pm$0.0146 & 507.0$\pm$48.0 & $-$0.848$\pm$0.036 & $-$2.175$\pm$0.082 & 531.91/476 \\
10.764 & 11.459 & 0.1997$\pm$0.0205 & 325.2$\pm$26.5 & $-$0.686$\pm$0.048 & $-$2.106$\pm$0.065 & 550.81/476 \\
11.459 & 12.253 & 0.1464$\pm$0.0132 & 415.6$\pm$37.5 & $-$0.808$\pm$0.040 & $-$2.154$\pm$0.081 & 539.34/476 \\
12.253 & 12.859 & 0.1434$\pm$0.0107 & 651.6$\pm$70.3 & $-$0.887$\pm$0.033 & $-$2.016$\pm$0.059 & 526.98/476 \\
12.859 & 13.366 & 0.2035$\pm$0.0156 & 588.3$\pm$64.5 & $-$0.859$\pm$0.036 & $-$1.931$\pm$0.046 & 640.58/476 \\
13.366 & 13.791 & 0.1857$\pm$0.0129 & 841.1$\pm$108.0 & $-$0.962$\pm$0.030 & $-$1.922$\pm$0.053 & 534.27/476 \\
13.791 & 14.244 & 0.1794$\pm$0.0163 & 582.8$\pm$95.4 & $-$0.992$\pm$0.039 & $-$1.749$\pm$0.038 & 514.99/476 \\
14.244 & 14.643 & 0.1721$\pm$0.0118 & 1066.0$\pm$188.0 & $-$1.131$\pm$0.027 & $-$1.880$\pm$0.057 & 512.53/476 \\
14.643 & 15.140 & 0.1476$\pm$0.0109 & 974.4$\pm$181.0 & $-$1.190$\pm$0.027 & $-$1.945$\pm$0.069 & 508.88/476 \\
15.140 & 15.815 & 0.1257$\pm$0.0093 & 839.6$\pm$146.0 & $-$1.279$\pm$0.025 & $-$2.154$\pm$0.111 & 556.16/476 \\
15.815 & 16.659 & 0.1009$\pm$0.0076 & 732.4$\pm$103.0 & $-$1.327$\pm$0.024 & $-$3.402$\pm$1.230 & 521.86/476 \\
16.659 & 17.717 & 0.0904$\pm$0.0094 & 409.8$\pm$58.2 & $-$1.402$\pm$0.029 & $-$2.917$\pm$0.751 & 546.90/476 \\
17.717 & 19.180 & 0.0648$\pm$0.0064 & 457.8$\pm$65.5 & $-$1.418$\pm$0.027 & $<-$4.091 & 536.98/476 \\
19.180 & 26.218 & 0.0566$\pm$0.0115 & 63.0$\pm$6.1 & $-$1.419$\pm$0.078 & $-$2.290$\pm$0.085 & 568.75/476 \\
26.218 & 36.930 & 0.0266$\pm$0.0042 & 124.2$\pm$18.0 & $-$1.316$\pm$0.062 & $-$2.010$\pm$0.068 & 725.95/476 \\
36.930 & 38.530 & 0.0848$\pm$0.0078 & 392.3$\pm$33.1 & $-$0.938$\pm$0.036 & $-$2.682$\pm$0.290 & 591.78/476 \\
38.530 & 40.034 & 0.1075$\pm$0.0106 & 330.0$\pm$25.9 & $-$0.821$\pm$0.042 & $-$2.380$\pm$0.128 & 547.79/476 \\
40.034 & 41.440 & 0.2016$\pm$0.0274 & 175.3$\pm$11.3 & $-$0.691$\pm$0.060 & $-$2.374$\pm$0.109 & 531.31/476 \\
41.440 & 43.824 & 0.1237$\pm$0.0176 & 165.6$\pm$9.2 & $-$0.858$\pm$0.049 & $-$2.774$\pm$0.222 & 446.06/476 \\
43.824 & 46.694 & 0.0996$\pm$0.0167 & 153.8$\pm$7.5 & $-$0.885$\pm$0.047 & $-$3.717$\pm$0.974 & 531.36/476 \\
46.694 & 49.917 & 0.1140$\pm$0.0179 & 130.0$\pm$7.6 & $-$0.648$\pm$0.067 & $-$2.360$\pm$0.111 & 571.19/476 \\
49.917 & 50.937 & 0.1577$\pm$0.0173 & 274.3$\pm$19.5 & $-$0.763$\pm$0.046 & $-$2.476$\pm$0.144 & 516.05/476 \\
50.937 & 51.685 & 0.2721$\pm$0.0325 & 214.4$\pm$13.0 & $-$0.534$\pm$0.055 & $-$2.285$\pm$0.085 & 497.29/476 \\
51.685 & 52.584 & 0.2729$\pm$0.0332 & 211.8$\pm$11.0 & $-$0.604$\pm$0.049 & $-$2.789$\pm$0.177 & 569.27/476 \\
52.584 & 54.244 & 0.1599$\pm$0.0212 & 186.7$\pm$10.7 & $-$0.708$\pm$0.053 & $-$2.617$\pm$0.162 & 574.81/476 \\
54.244 & 56.959 & 0.1294$\pm$0.0185 & 156.2$\pm$8.9 & $-$0.739$\pm$0.057 & $-$2.570$\pm$0.161 & 579.22/476 \\
56.959 & 77.824 & 0.0809$\pm$0.0246 & 63.0$\pm$4.2 & $-$0.590$\pm$0.129 & $-$2.309$\pm$0.072 & 1006.3/476 \\
\hline
\end{tabular}
\end{table*}

\clearpage

\begin{figure*}[!htbp]
\resizebox{\hsize}{!}
{\includegraphics[width = 18 cm]{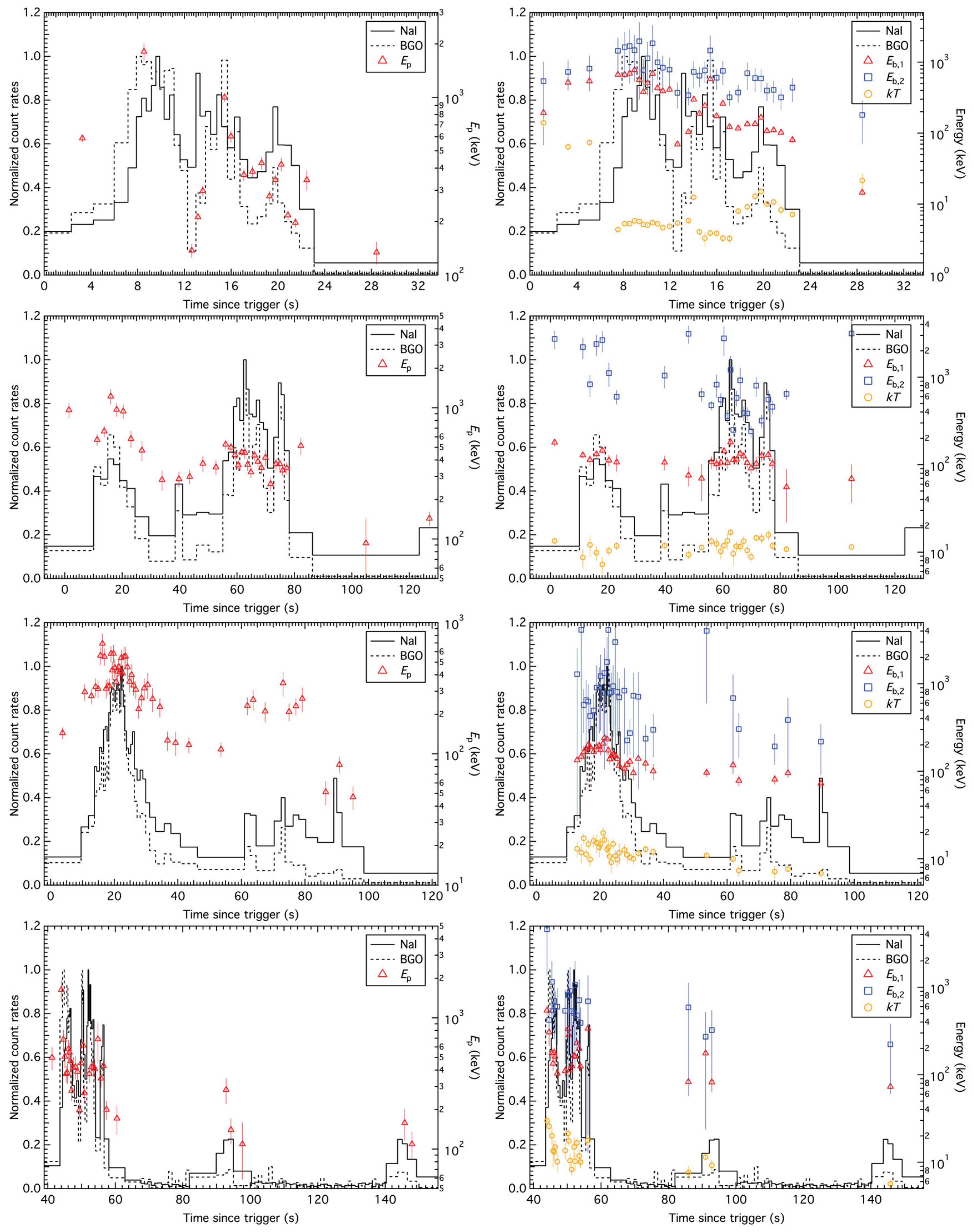}}
\caption{Panels from top to bottom: light curves of GRB 090902B, GRB 100724B, GRB 100826A, and GRB 101123A with the evolutions of constrained $E_\text{p}$, $E_\text{b,1}$, $E_\text{b,2}$, and $kT$ overlaid.}
\label{fig:AppFig1}
\end{figure*}

\clearpage

\begin{figure*}[!htbp]
\resizebox{\hsize}{!}
{\includegraphics[width = 18 cm]{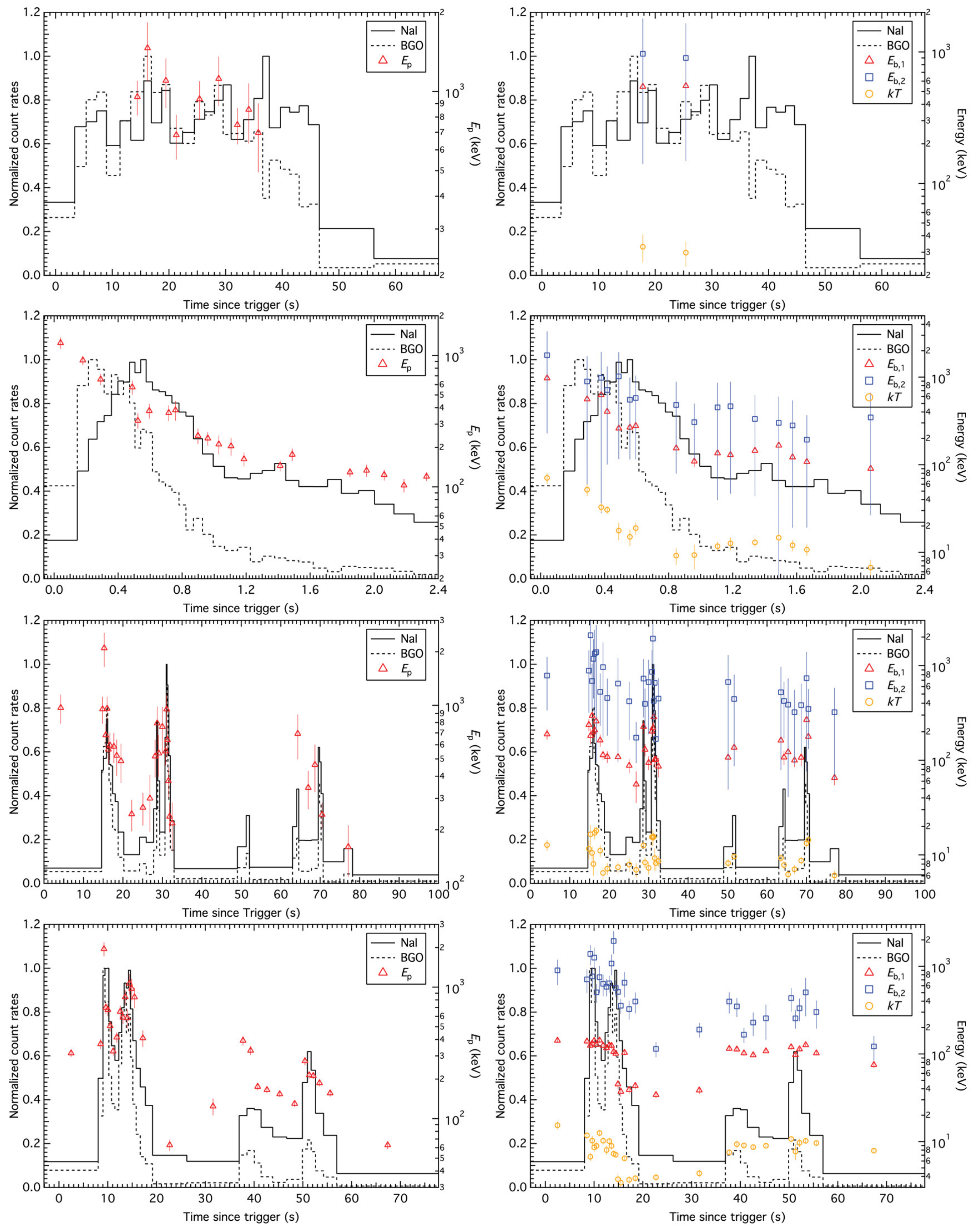}}
\caption{Panels from top to bottom: light curves of GRB 120526A, GRB 130427A, GRB 130504C, and GRB 130606B with the evolutions of constrained $E_\text{p}$, $E_\text{b,1}$, $E_\text{b,2}$, and $kT$ overlaid.}
\label{fig:AppFig2}
\end{figure*}

\clearpage

\end{appendix}

\end{document}